\documentclass[fleqn,usenatbib]{mnras}

\usepackage{newtxtext,newtxmath}

\usepackage[T1]{fontenc}

\DeclareRobustCommand{\VAN}[3]{#2}
\let\VANthebibliography\thebibliography
\def\thebibliography{\DeclareRobustCommand{\VAN}[3]{##3}\VANthebibliography}

\usepackage{graphicx}	
\usepackage{amsmath}	
\usepackage{hyperref}
\usepackage{wasysym}
\usepackage{multirow}
\usepackage{array}
\usepackage{subfigure}
\usepackage{placeins}
\usepackage{nccmath}
\extrafloats{500}

\newcommand{\solarm}{\ensuremath{M_\odot}}
\newcommand{\kms}{\ifmmode\,{\rm km}\,{\rm s}^{-1}\else km$\,$s$^{-1}$\fi}
\newcommand{\Rd}{\ifmmode\,R_{\rm d}\else $R_{\rm d}$\fi}
\newcommand{\be}{\begin{equation}}
\newcommand{\ee}{\end{equation}}
\newcommand\lt{<}
\newcommand\ltsima{$\; \buildrel < \over \sim \;$}
\newcommand\ltsim{\lower.5ex\hbox{\ltsima}}
\newcommand\gtsima{$\; \buildrel > \over \sim \;$}
\newcommand\gtsim{\lower.5ex\hbox{\gtsima}}
\newcommand{\magarc}{\ifmmode {{{{\rm mag}~{\rm arcsec}}^{-2}}}
             \else {{{mag}$~${arcsec}$^{-2}$}}
             \fi}

\def \ion#1#2{#1{\footnotesize{#2}}\relax}
\newcommand{\othree}{[O~\textsc{iii}]}
\newcommand{\oone}{[O~\textsc{i}]}

\newcommand{\ntwo}{[N~\textsc{ii}]}

\newcommand{\hetwo}{He~\textsc{ii}}
\newcommand{\heone}{He~\textsc{i}}

\def \ha {H$\alpha$}
\def \hi{\ion{H}{I}\ }

\newcommand{\atlas}{\textsc{ATLAS}$^{\rm 3D}$\ }

\newcommand\pPXF{\texttt{pPXF}\ }

\def\se#1{\S\ref{sec:#1}}
\def\Eq#1{Eq.~(\ref{eq:#1})}

\def\Fig#1{Fig.~\ref{fig:#1}}
\def\Table#1{Table~\ref{tab:#1}}

\title[SHIVir Survey: Data Catalogue]{The Spectroscopy and H-band Imaging of Virgo cluster galaxies (SHIVir) Survey: Data Catalogue and Kinematic Profiles}

\author[N. N.-Q. Ouellette et al.]{Nathalie N.-Q. Ouellette$^{1}$\thanks{E-mail: nathalie@astro.umontreal.ca},
St\'ephane Courteau$^{2},$
Jon A. Holtzman$^{3},$
Michael McDonald$^{4},$
\newauthor
Michele Cappellari$^{5},$
Joel C. Roediger$^{6},$
Patrick C\^{o}t\'{e}$^{6},$
Julianne J. Dalcanton$^{7},$
\newauthor
Elena Dalla Bont\`{a}$^{8,9},$
Laura Ferrarese$^{6},$
R. Brent Tully$^{10},$
Connor Stone$^{2},$ 
Eric W. Peng$^{11,12}$
\\
$^{1}$Institute for Research on Exoplanets, Department of Physics, Universit\'{e} de Montr\'{e}al, Montr\'{e}al, QC, H3C 3J7, Canada\\
$^{2}$Department of Physics, Engineering Physics and Astronomy, Queen's University, Kingston, ON K7L 3N6, Canada\\
$^{3}$Department of Physics and Astronomy, New Mexico State University, Las Cruces, NM, 88003-8001, USA\\
$^{4}$MIT Kavli Institute for Astrophysics and Space Research, MIT, Cambridge, MA, 02139, USA\\
$^{5}$Sub-department of Astrophysics, Department of Physics, University of Oxford, Denys Wilkinson Building, Keble Road, Oxford, OX1 3RH, UK\\
$^{6}$Herzberg Astronomy and Astrophysics Research Centre, National Research Council of Canada, 5071 W. Saanich Road, Victoria, BC, V9E 2E7, Canada\\
$^{7}$Department of Astronomy, University of Washington, Seattle, WA, 98195, USA\\
$^{8}$Dipartimento di Fisica e Astronomia ``G. Galilei'', Universit\`{a} di Padova, vicolo dell'Osservatorio 3, I-35122, Padova, Italy\\
$^{9}$INAF --- Osservatorio Astronomino di Padova, vicolo dell'Osservatorio 5, I-35122, Padova, Italy\\
$^{10}$Institute for Astronomy, University of Hawaii, 2680 Woodlawn Drive, Honolulu, HI 96822-1839, USA\\
$^{11}$Department of Astronomy, Peking University, Beijing 100871, China\\
$^{12}$Kavli Institute for Astronomy and Astrophysics, Peking University, Beijing 100871, China
}

\date{Accepted 2022 May 10. Received 2022 May 8; in original form 2021 May 2.}

\pubyear{2022}

\begin{document}
\label{firstpage}
\pagerange{\pageref{firstpage}--\pageref{lastpage}}
\maketitle

\begin{abstract}
The ``Spectroscopy and H-band Imaging of Virgo cluster galaxies'' (SHIVir) survey is an optical and near-infrared survey which combines SDSS photometry, deep H-band photometry, and long-slit optical spectroscopy for 190 Virgo cluster galaxies (VCGs) covering all morphological types over the stellar mass range $\log (M_*/\solarm) = 7.8-11.5$.  
We present the spectroscopic sample selection, data reduction, and analysis for this SHIVir sample. 
We have used and optimised the \texttt{pPXF} routine to extract stellar kinematics from our data.
Ultimately, resolved kinematic profiles (rotation curves and velocity dispersion profiles) are available for 133 SHIVir galaxies. 
A comprehensive database of photometric and kinematic parameters for the SHIVir sample is presented with: grizH magnitudes, effective surface brightnesses, effective and isophotal radii, rotational velocities, velocity dispersions, and stellar and dynamical masses. 
Parameter distributions highlight some bimodal distributions and possible sample biases. 
A qualitative study of resolved extended velocity dispersion profiles suggests a link between the so-called ``sigma-drop'' kinematic profile and the presence of rings in lenticular S0 galaxies. 
Rising dispersion profiles are linked to early-type spirals or dwarf ellipticals for which a rotational component is significant, whereas peaked profiles are tied to featureless giant ellipticals.
\end{abstract}




\section*{Note}

This is a pre-copyedited, author-produced PDF of an article accepted for publication in /mnras following peer review. The version of record [Nathalie N-Q Ouellette, Stéphane Courteau, Jon A Holtzman, Michael McDonald, Michele Cappellari, Joel C Roediger, Patrick Côté, Julianne J Dalcanton, Elena Dalla Bontà, Laura Ferrarese, R Brent Tully, Connor Stone, Eric W Peng, The spectroscopy and H-band imaging of Virgo cluster galaxies (SHIVir) survey: data catalogue and kinematic profiles, Monthly Notices of the Royal Astronomical Society, Volume 514, Issue 2, August 2022, Pages 2356–2375. doi: 10.1093/mnras/stac1347] is available online at: \href{https://academic.oup.com/mnras/advance-article-abstract/doi/10.1093/mnras/stac1347/6608887}{https://academic.oup.com/mnras/advance-article-abstract/doi/10.1093/mnras/stac1347/6608887}.

\section{Introduction}
\label{sec:Intro}

A global understanding of galaxy formation and evolution requires that the broadest range of structural and environmental galaxy parameters including shapes, sizes, mass fractions, activity levels, environmental densities be reproduced by models. 
Deep and extensive surveys providing high-quality spatially-resolved digital spectrophotometric data for thousands of galaxies are needed to understand the physical drivers of galaxy formation and evolution in a statistical manner. 
Surveys such as SAURON \citep{Bacon2001}, SDSS \citep{Abazajian2003}, ATLAS$^{3{\rm D}}$ \citep{Cappellari2011a}, GAMA \citep{Robotham2011}, CALIFA \citep{Sanchez2012}, SAMI \citep{Croom2012}, MaNGA \citep{Bundy2015}, and others have provided valuable early explorations. 
Integral field and longslit spectroscopy techniques are typically used to extract chemical and kinematic information for galaxy samples ranging from a few hundred (e.g., SAURON) to several thousand (e.g., MaNGA) targets.

Of particular value in contrasting observations with simulations and/or theoretical predictions are tracers of galaxy structure that enable some discrimination between plausible models. 
High resolution multi-band data and extended light and rotation profiles are, of course, especially desirable for such analyses. 
In this context, the ``Spectroscopy and H-band Imaging of Virgo cluster galaxies'' (SHIVir) survey, which brings together optical/near-infrared photometry and spectroscopy for a large sample of Virgo cluster galaxies (VCGs), is a unique addition to the growing list of public catalogues of galaxy parameters. 
SHIVir offers a large set of well-resolved and spatially-extended kinematic and photometric properties of late-type and early-type galaxies (LTGs/ETGs, respectively) belonging to the Virgo cluster. 
While SHIVir's galaxy catalogue may be smaller than those of large surveys such as MaNGA \citep{Bundy2015}, its extended kinematics probing out to the transition from baryon- to dark-matter-dominated regions sets it apart and proves to be a most complementary database for the study of global galaxy properties.

Contrasting with the integral field spectroscopy (IFS) used in recent spectroscopic surveys of galaxies (see above), the long-slit spectroscopy presented in SHIVir provides a spatially extended sampling of galaxy structure due to its superior radial coverage. 
This is especially true for low-mass ETGs for which few resolved kinematic studies exist, given their faint outskirts.  
Current integral field unit (IFU) velocity maps are typically limited to one effective radius $R_{\rm e}$ (at the distance of Virgo), thwarting the measurement of important dynamical parameters at larger radii~\citep{Courteau1997,Cappellari2006,Courteau2007,Cappellari2013a}. 
SHIVir's use of long-slit spectroscopy coupled with long exposure times is largely motivated by the desire to probe the crucial baryon-to-dark-matter dominated transition regions in galaxies between 1 and 4 $R_{\rm e}$ \citep{Courteau2014}.

A dynamically active, structurally complex, and nearby rich galaxy cluster, the Virgo cluster is an invaluable yet challenging study subject. 
Its proximity and richness make it a superb laboratory for measuring galaxy structural properties (sizes, velocities, colours, gas masses, etc.) and characterising IMF variations and star formation rates within and among galaxies. 
Being a cosmologically-representative sample of local galaxies \citep{Binggeli1987}, the Virgo cluster is an ideal testing ground to study evolutionary signatures in more distant systems and remains the subject of intense scrutiny. 
More detailed explanations for the choice of the Virgo cluster as SHIVir's main target can be found in \citet[hereafter O17]{Ouellette2017}. As a complement to other extensive investigations of the Virgo cluster \citep{Cote2004, Chung2009, Davies2010, Boselli2011, Ferrarese2012, Ferrarese2020}, SHIVir offers a unique combination of optical and infrared wide-field imaging as well as spatially-extended optical long-slit spectroscopy for a wide areal coverage of the Virgo cluster. 
This offers a broad census of galaxy stellar masses, ages and metallicities \citep{Roediger2011a,Roediger2011b,Roediger2015}. 

\subsection{Paper Layout}

The main goal of this paper is to present and make available the SHIVir spectroscopic sample. 
An introduction to the overall SHIVir survey, including its conception, goals and photometric complement, is found in \se{SHIVir_Intro}, followed by a description of SHIVir's spectroscopic sample in \se{SHIVir_Spec}. 
The data collection and spectroscopy reduction methods are described in \se{Reduction}, and the kinematic measurements and inferred dynamical masses are presented in \se{DataAnalysis}. 
The latter section also includes a systematic analysis of velocity dispersion (VD) extraction using \citet{Cappellari2017}'s \texttt{pPXF} routine and the biases incurred due to its algorithmic configuration. 
\se{param_dist} presents parameter distributions of the galaxy population of the spectroscopic SHIVir sample (SpecSHIVir) of 190 galaxies, and \se{kinprofiles} presents all LTG rotation curves (RCs) and ETG VD profiles along with a proposed classification for the latter.
Conclusions and a discussion of future explorations based on the SHIVir project are presented in \se{Conclusion}. 
All available resolved kinematic profiles (\ha\ rotation curves and VD profiles) are presented in the paper's online supplementary material (see Data Availability), along with notes for each SHIVir galaxy for which such profiles could be extracted. 
The first few rows of the following data tables are also presented in the Data Availability section, the entirety of which being available in the supplementary material: (i) a target list of all SpecSHIVir galaxies with basic photometric and observational characteristics, (ii) kinematic data including rotational velocities and velocity dispersions, and (iii) a table of circular velocities and masses.


\section{Survey Presentation}
\label{sec:SHIVir_Intro}

The SHIVir survey was first introduced in \citet[][hereafter M09]{McDonald2009b} to investigate a possible surface brightness (SB) bimodality in spiral galaxies \citep{Tully1997}. 
Extending beyond the SB bimodality in early and late-type galaxies, our (then) largest photometric and spectroscopic database for a single galaxy cluster was assembled in order to address various outstanding issues in galaxy formation and evolution in a cluster environment. 
A summary of SHIVir's subsequent developments was presented in O17.

In this section, we discuss the selection of VCGs for the SHIVir survey and give an overview of the SHIVir photometry and its derived parameters. 
The SHIVir spectroscopy and its derived parameters are presented in \se{SHIVir_Spec} -- \se{DataAnalysis}.

\subsection{Catalogue}\label{sec:Catalog}

\begin{table}
\begin{center}
{\small
\begin{tabular}{c|c}
\hline\hline
Virgo Set & Number of galaxies\\
\hline									
VCC & 2095\\
VCC/SDSS & 742\\
H-band & 286\\
SpecSHIVir & 190\\
\hline\hline
\end{tabular} }
\caption{The number of VCGs included in each subset described in this paper. 
(1) Galaxies included in \citet{Binggeli1985}'s Virgo Cluster Catalogue (VCC). 
(2) Number of VCC galaxies with available $i$-band photometry in the SDSS DR6 database. 
(3) Number of VCC/SDSS galaxies for which $H$-band photometry was obtained by the SHIVir team. 
(4) Number of VCC/SDSS galaxies for which longslit spectroscopy was obtained by the SHIVir team.}
\label{tab:catalog_info}
\end{center}
\end{table}

\begin{figure*}
\centering
\includegraphics[width=0.99\textwidth]{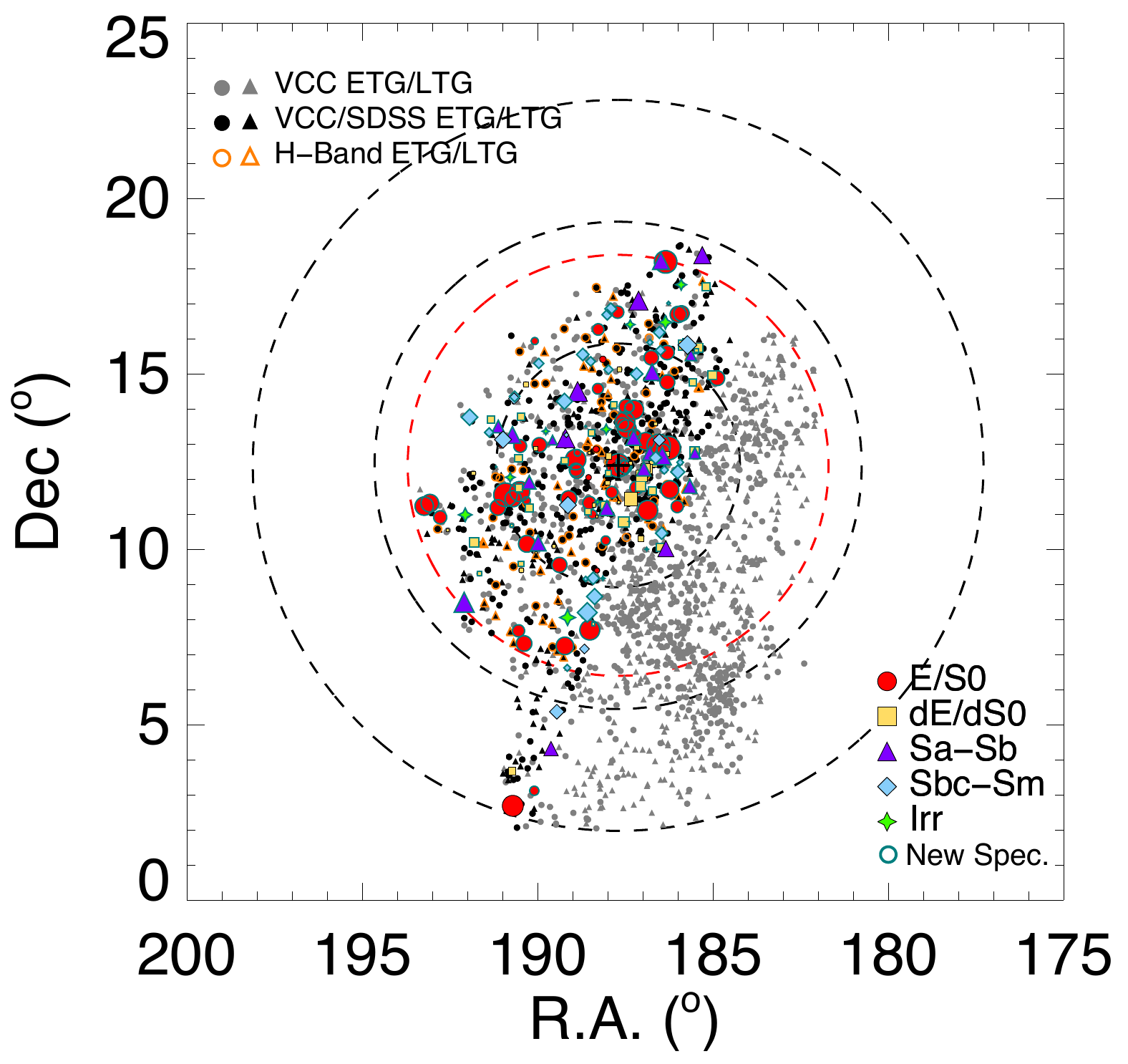}
\caption{Two-dimensional map of the Virgo cluster of galaxies. 
Black points are galaxies in our VCC/SDSS sample (M09) while grey points are the remaining galaxies in the full VCC \citep{Binggeli1985}. 
These are labeled as ETGs (circles) and LTGs (triangles). 
Galaxies with available $H$-band photometry (M09) are highlighted with an orange border. 
Coloured points are coded by morphology for galaxies that belong to our spectroscopic SHIVir sample (SpecSHIVir, 190 galaxies). 
The size of these points scales with total luminosity. 
Galaxies for which we obtained new spectroscopy are highlighted with a teal border. 
The concentric dashed circles are projected galactocentric distances (1, 2 and 3 Mpc) around the centre of the Virgo cluster at M87 (marked by a black cross). 
The dashed red circle corresponds to a distance of $6^{\circ}$ from M87, as defined by M09.}
\label{fig:SHIVir_Map}
\end{figure*}

The SHIVir survey draws its sample from the magnitude-limited Virgo Cluster Catalog \citep[VCC,][]{Binggeli1985} containing 2096 galaxies in a 140 deg$^{2}$ area around the central galaxies M49 and M87 at $\alpha{\sim}12^{\rm h}25^{\rm m}$ and $\delta{\sim}13^{\circ}$. 
The spatial distribution of all VCC galaxies, including members of the subsamples described in this paper, is shown in \Fig{SHIVir_Map}. 

\begin{figure*}
\centering
\includegraphics[width=0.8\textwidth]{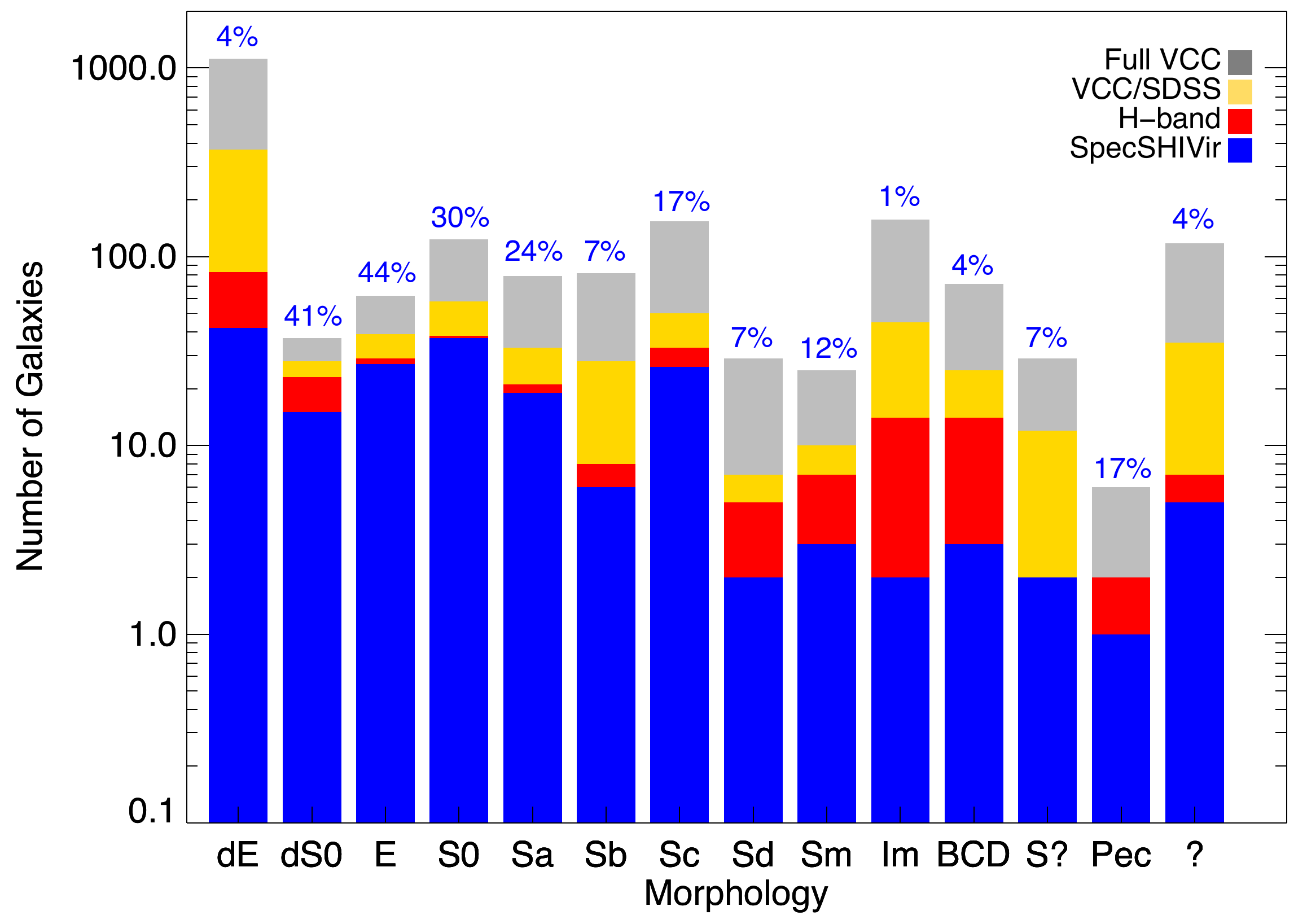}
\caption{Distribution of galaxy morphologies for the full VCC catalogue (gray), the VCC/SDSS subsample (yellow), the $H$-band subsample (red), and the spectroscopic SHIVir (SpecSHIVir) subsample (blue). 
The fraction of the full VCC catalogue covered by the SpecSHIVir subsample for each morphological bin is shown in blue above each histogram bin. 
The morphological classification is taken from the GOLDMine database \citep{Gavazzi2003}.} 
\label{fig:SHIVir_Morphology}
\end{figure*}

The spatial cut to reject the W, W', and M background groups \citep{deVaucouleurs1961,Ftaclas1984}, shown to be 1.5 to 2 times more distant than the Virgo cluster itself, is described in Section 6.1 \cite[see Figures 9 and 10 of][]{Tully2016}.
With these contaminants discarded, the full SHIVir sample (referred to as ``VCC/SDSS'' in \Fig{SHIVir_Map} and \Fig{SHIVir_Morphology}) contains 742 VCGs for which $g$-, $r$- and $i$-band images were available in the SDSS 6th Data Release \citep{AdelmanMcCarthy2008} when M09 assembled the original photometric component of SHIVir.
More recent $g$-, $r$, and $z$ photometry has also been extracted from the Dark Energy Spectroscopic Instrument Legacy Survey \citep{Dey2019}.
These data are reported in \citet{Stone2021a}. 

\begin{figure*}
\centering
\includegraphics[width=0.8\textwidth]{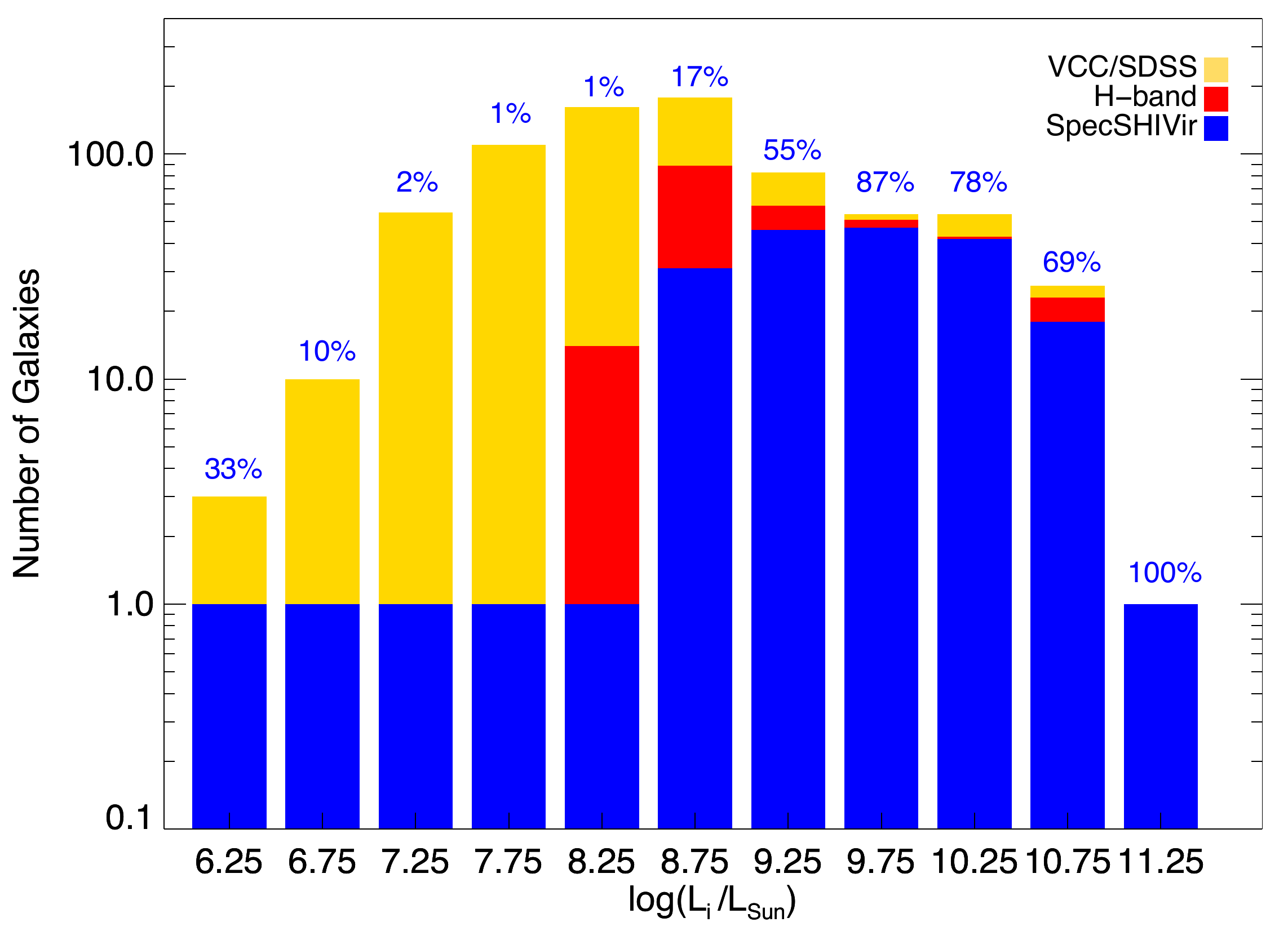}
\caption{Distribution of galaxy total $i$-band luminosities in the VCC/SDSS sample (yellow), the $H$-band subsample (red), and the spectroscopic SHIVir subsample (blue). 
The fraction of the full VCC catalogue covered by the SpecSHIVir subsample for each luminosity bin is included in blue above each histogram bin.} 
\label{fig:SHIVir_Luminosity}
\end{figure*}

A representative subsample (referred to as the ``H-band'' sample in \Fig{SHIVir_Map} and \Fig{SHIVir_Morphology}) taken from the VCC/SDSS sample was imaged in the $H$-band as part of the SHIVir survey (M09) in order to investigate SB bimodality in Virgo cluster galaxies.
This required a nearly complete sample down to at least $M_{B} = -15.15$ mag, in order to trace the low-SB peak found for the Ursa Major cluster in \citet{Tully1997}. 
Two additional spatial cuts were performed: (1) galaxies identified as `background' in the VCC, and (2) galaxies over $6^{\circ}$ away from M87, defined as Virgo's centre, were removed (see \Fig{SHIVir_Map}, dashed red circle, for this delimitation). 
Eight VCC galaxies were removed for having recessional velocities greater than 3000 km s$^{-1}$: VCC 1068, VCC 1217, VCC 1258, VCC 1355, VCC 1665, VCC 1768, VCC 1889, and VCC 2096. VCC 723 and VCC 991 were discarded due to significant foreground star contamination. Seven VCC galaxies could not be detected in the $H$-band: VCC 530, VCC 950, VCC 1052, VCC 1287, VCC 1571, VCC 1822, and VCC 1992. 

The final $H$-band SHIVir catalogue contains 286 VCC galaxies covering a range of morphologies and luminosities. 
This limits any biases due to selection effects such as morphological segregation. 
\Fig{SHIVir_Morphology} shows that all morphological types are fairly represented in both the full SHIVir sample (VCC/SDSS, yellow bars) and the $H$-band SHIVir sample (red bars), when compared to the full VCC (gray bars). 
A higher proportion of ETGs have been observed relative to LTGs; SHIVir also suffers from a dearth of dwarf galaxies (especially dEs) due to our magnitude limit, but M09 shows that their conclusions are still robust in light of this limitation. 
The dearth of dwarfs reflects the challenge of observing these galaxies spectroscopically and is particularly noticeable in our SpecSHIVir subsample (blue bars; the SpecSHIVir subsample is described in \se{SHIVir_Spec}). 
The low SNR is a real concern for the extraction of kinematics which depends on absorption features. 
See \Table{catalog_info} for the number of galaxies in each VCG subset included in this paper. 

Along similar lines, we have included a distribution of the total $i$-band luminosities of our different SHIVir subsamples in \Fig{SHIVir_Luminosity}. 
Here, the higher luminosity bins are closer to completeness. 
An important future goal is to augment the SHIVir catalogue with low-luminosity/mass systems for extended studies of Virgo cluster galaxy scaling relations such as those shown in O17.

\subsection{Distances}
\label{sec:distances}

If available, distances to individual VCGs were taken from \citet{Mei2007} and \citet{Blakeslee2009}; otherwise, a distance of 16.5 Mpc \citep{Mei2007} was assumed. 
The distance of M87, used to calculate the physical galactocentric radii of Virgo cluster members, was taken to be 16.5 Mpc \citep{Blakeslee2009}. 
Distances were not available for the LTG population of the SpecSHIVir subsample. 
The issue of morphological segregation, whereby LTGs are more likely to populate the outskirts of a cluster, and the effect of distance uncertainty on our scaling relations and mass relations, are addressed in O17.

\subsection{Photometry}
\label{sec:Photometry}

Numerous studies of galaxy structure and evolution, and comparisons with numerical models, rely on matching photometric and spectroscopic data for galaxies.  
This is especially relevant in the context of galaxy scaling and parametric relations.
Below we review the SDSS photometry used for the VCC/SDSS SHIVir sample.
Spectroscopy will be discussed in \se{SHIVir_Spec}.

Two separate sets of photometric profiles make up our SHIVir catalog: one led by Michael McDonald (hereafter MM, published in M09), and the other led by Yucong Zhu (hereafter YZ). 
MM and YZ used $g$, $r$ and $i$-band images from the SDSS DR6 catalogue \citep{AdelmanMcCarthy2008} for 286 and 742 Virgo galaxies respectively (catalogue selection was described in \se{Catalog}), and extracted SB profiles and total luminosities via isophotal fitting. 
MM also analysed $H$-band images for their subset of 286 VCGs using publicly available Two Micron All-Sky Survey \citep[20 VCGs]{Skrutskie2006} and GOLDMine \citep[79 VCGs]{Gavazzi2003} imaging, as well as new photometry collected with ULBCAM at the University of Hawai'i 2.2-m telescope (122 VCGs), WFCAM at the United Kingston InfraRed Telescope (31 VCGs), and WIRCAM at the Canada-France-Hawai'i Telescope (34 VCGs) between 2005 and 2008. 
Details about the photometry reduction and analysis are given in M09.

Unless otherwise noted, $i$-band images and derived parametric quantities are used for structural correlations since this redder band is less sensitive to selective dust extinction \citep{Peletier1992,deJong1996}. 
Because our dynamical study of the Virgo cluster includes galaxies taken from the larger SHIVir catalogue of 742 galaxies, we use the parameters measured by YZ to homogenise our photometric catalogue. 
Both YZ and MM's inferred parameters are considered whenever possible.

Galaxy light profiles can be modelled either by imposing a multi-component decomposition (parametric approach) to account for the different structures present such as the disc and bulge \citep{MacArthur2003,McDonald2009a}, or via a non-parametric approach, free of model assumptions \citep{Stone2021b}. 
M09 demonstrated, however, that their conclusions about SB bimodality are robust to the analysis method used. 
Since the VCGs used in our spectroscopic analysis cover a wide range of galaxy morphologies that themselves contain an array of complex structures (spiral arms, warped discs, compact nuclei, etc.), we prefer photometric quantities derived in non-parametric fashion.
The only two assumptions made are that: 1) the total galaxy light is an extrapolation of the light profile to infinity (see footnote 2), and 2) inclination estimates and extinction corrections are valid (M09). 
Light profiles were created by performing isophotal ellipse fitting on the images, where elliptical contours passing through specific SB levels were drawn on the galaxy image. Relevant photometric quantities derived by MM and YZ include: $C_{28}$ concentration\footnote{A higher $C_{28}$ measurement indicates that a galaxy is more centrally concentrated, which is characteristic of ETGs with large bulges. 
LTGs with small bulges have lower $C_{28}$ values. 
A S\'ersic $n=1$ profile has $C_{28}=2.8$ whereas a S\'ersic $n=4$ profile has $C_{28}=5.1 ~$\citep{MacArthur2003,Vitral2021}.}, defined as

\begin{equation}
\label{eq:c28}
C_{28} = 5 {\rm log} \bigg{(} \frac{r_{80}}{r_{20}} \bigg{)},
\end{equation}

\noindent where $r_{80}$ and $r_{20}$ are the radii enclosing 80\% and 20\% of the total galaxy light respectively, isophotal radii, disc scale length, effective SB $\mu_{e}$, effective radius $R_{\rm e}$ and total luminosity\footnote{$L_{\rm tot}$ and $R_{\rm e}$ may rely on model-dependent extrapolation of the light profile (growth curve) of a galaxy to infinity or the definition of a galaxy's edge.}.

Position angle is the angle of a galaxy's semi-major axis relative to the north celestial pole. 
The ellipticity $\epsilon = (1 - \frac{b}{a})$ of these outermost isophotes, where $a$ and $b$ are the semi-major and semi-minor axes, is also used to determine the approximate inclination $i$ of disc galaxies using Hubble's oblate spheroid description \citep{Hubble1926}:

\begin{equation}
{\rm cos\ } i = \sqrt{ \frac{(1 - \epsilon)^{2} - q^{2}}{1 - q^{2}} },
\end{equation}

\noindent where $q$ is the intrinsic axial ratio of the spiral disc equal to $c/a$, where $c$ is the polar axis \citep{Giovanelli1997}. \citet{Hall2012} studied the variations of these axial ratios in the $i$-band versus mean galaxy colour for a sample of 871 galaxies, and found $q \approx 0.13$ to be the minimum axial ratio and thus the intrinsic thickness of spiral galaxies, in agreement with \citet{Giovanelli1994}. 
This axial ratio was adopted to compute disc inclinations.

SB, isophotal radius and effective radius were all corrected for inclination. 
In addition to this geometric correction, all luminosity, magnitude and SB values must be corrected for wavelength-dependent dust extinction. 
For SB, we use the geometric and extinction terms presented by \citet{Tully1997} and \citet{Tully1998}:

\begin{equation}
\mu^{\rm i} = \mu - 2.5 C^{\lambda} {\rm log} (b/a),
\end{equation}

\noindent where $\mu^{\rm i}$ is the corrected SB, $b$ and $a$ are the semi-minor and semi-major axes, and $C^{\lambda}$ is a wavelength dependent extinction correction. 
For low SB galaxies ($\mu_{\rm e} > 21$ \magarc), we used $C^{i} = 1$ for our $i$-band photometry. 
For high SB galaxies ($\mu_{\rm e} < 21$ \magarc), we used $C^{i} = 0.61$. 
All magnitudes were corrected for dust attenuation using the extinction terms from \citet{Schlafly2011}.

Total apparent $i$-band magnitude, effective SB, effective radius, 23.5 \magarc isophotal radius, and inclination for all SHIVir VCGs featured in this work (190 galaxies included in the SpecSHIVir subsample described in \se{SHIVir_Spec}) can be found in our descriptive target list in the supplementary material (see \Table{target_list}). 
The SHIVir photometric catalogue was introduced in M09 ($H$-band imaging) and \citet[][multi-band imaging and stellar population analysis]{Roediger2011a,Roediger2011b}; refer to these for a full description of the photometric analysis. 
The M09 data are available at \texttt{https://www.queensu.ca/academia/courteau/links}.

\subsection{Stellar Masses}
\label{sec:StellarMasses}

While the topic of dynamical/total masses is discussed in \se{DataAnalysis}, since spectra are required for their measurement, our stellar masses rely strictly on photometry and stellar populations. 
Our stellar mass estimates exploit colour transformations such as those presented in \citet{Roediger2015}. 
SDSS colours $g-r$, $g-i$, $g-z$, and $g-H$ were used to constrain mass-to-light versus colour relations (MLCRs). 
This technique allows for the straightforward modelling and fitting of spectral energy distributions (SEDs) using libraries of stellar templates and assuming for reference a universal Chabrier IMF \citep{Chabrier2003}, which then yield stellar mass-to-light ratios from which a stellar mass can be inferred.
Stellar masses for SHIVir galaxies inferred using this technique were presented in O17.

Colours are computed from the SHIVir photometric catalogue described above. 
The stellar mass errors presented here account for random uncertainties. 
Additional systematic errors amounting to 0.3~dex may be due to the choice of IMF
and other modeling decisions \citep{Conroy2013,Courteau2014,Roediger2015}.
Such systematic errors also directly impact the mass assessment of a putative dark matter component calculated as the total mass minus the stellar (baryonic) mass (see \se{dyn_mass}).


\section{Spectroscopic Sample Selection}
\label{sec:SHIVir_Spec}

The detailed study of Virgo cluster galaxies requires a complete set of photometric and spectroscopic parameters. 
The SHIVir team collected long-slit spectroscopy to measure spatially-resolved galaxy kinematics within the optical radius of a galaxy, i.e. beyond the transition from baryon-to-dark matter dominance (1 to 4 $R_{\rm e}$) in galaxies. 
Because long-slit spectroscopy of small, diffuse galaxies, as found in the Virgo cluster, requires long integrations, our SHIVir sample is grossly incomplete at low luminosities, as discussed below. 

A complete list of the galaxies in the SpecSHIVir catalogue (190 VCGs), along with their NGC/UGC/IC designation (if available), $i$-band apparent magnitude, $i$-band effective SB, effective radius, 23.5 \magarc level isophotal radius, morphological type, distance, inclination, and telescope used for observation (if part of our novel spectroscopic SHIVir catalog) can be found in the ``Target List'' table in the supplementary material, the first few rows of which are reprinted in \Table{target_list}. 
The morphological classification taken from the GOLDMine database \citep{Gavazzi2003} corresponds to numerical Hubble types ranging from $-3$ to $20$. 
We treat galaxies with a Hubble type between $-3$ and $2$ inclusively as ETGs, and between $3$ and $20$ as LTGs.

\subsection{Sample Selection}

Spectroscopic data should ideally be available for the entire 286 galaxy $H$-band SHIVir subsample or, better yet, the entire 742 galaxy SHIVir catalogue. 
However our scientific goals, described in \se{SHIVir_Intro} and presented in O17, call for long-slit spectra with prohibitively long integration times to reach sufficient SNR levels. 
Consequently, a subsample of the $H$-band SHIVir catalogue was created to define our spectroscopic catalogue. 
While the latter is a fair representation of the overall Virgo population, it is not cosmologically representative, and many faint dwarf galaxies were missed given the required long integration times (i.e., multiple hours on 10-m class telescopes). \Fig{SHIVir_Morphology} and \Fig{SHIVir_Luminosity} show clearly that the fraction of dwarf galaxies (dE, Im, ?, etc.) in the spectroscopy sample, that we call ``SpecSHIVir'' is indeed quite low.
Galaxies with a major axis exceeding the slit length were also not selected, since some amount of blank sky along the slit is required for sky subtraction. 

Since \ha\ emission lines were used to determine LTG kinematics, a minimum galaxy inclination of 20$^\circ$ was adopted to ensure a sufficient Doppler shift component in the lines.
LTGs with inclinations greater than 70$^\circ$ were also excluded to minimise the effect of thick dust lanes on the projected galaxy disc.
For ETGs, high S/N absorption features are required to determine velocity dispersions, and galaxies with significant central emission (e.g., AGN activity, Seyfert types, etc. as categorised by the NASA Extragalactic Database) were not selected. All suitable LTGs and all ETGs with $m_{B} < 13.5$ ($M_{i} < -19.0$) were observed during our observing campaigns from 2008 to 2012. 
Subsequent campaigns in 2014-2015 focused on fainter dwarf ETGs to sample the lower-mass end (with $M_{*} < 10^{9} \solarm$) of the SHIVir catalogue.  
We also observed galaxies previously targeted by IFU surveys to compare our respective dispersions and study systematic differences between 2D and 3D spectroscopy. 
The faint dwarf selection ensured that each mass bin is populated by the same number of galaxies, working down in stellar mass within each mass bin. 
Re-observations favoured those with published profiles, for comparison, such as those observed by the ATLAS$^{3{\rm D}}$ collaboration \citep{Cappellari2011a}.

\subsection{Observations}

\begin{table*}
\begin{center}
{\small
\begin{tabular}{c||c|c|c}
\hline\hline
									    & KPNO							& APO (b/r)						    & Gemini \\
\hline									
Telescope								& Mayall 4.0m 					& ARC 3.5m 						    & Gemini South 8.2m \\
Detector								& T2KB (2048x2048) 				& e2v CCD42-40 (2048x1024) 		    & Hamamatsu (6266x4176) \\
Read Noise ($e^{-}$)					& 4.0 							& 4.9 / 4.6						    & 3.98 \\
Gain	($e^{-}$/ADU)					& 1.0 							& 1.68 / 1.88 					    & 1.83 \\
Spectrograph							& RC Spectrograph 					& DIS 						    & GMOS-S\\
Grating ($\ell$/mm)						& 632 							& B1200/R1200 					    & B1200 \\
Blaze (1$^{\rm st}$ order) (\r{A})	    & 5500 							& Not Available 					& 4630 \\
Spectral Range (\r{A})			        & 3925-5430 					& 4160-5420 / 6015-7200 		    & 4100--5500 \\
Slit Dimensions							& 1.5$\arcsec$ x 5.2$^{\prime}$		& 1.5$\arcsec$ x 6$^{\prime}$ 	& 1.0$\arcsec$ x 5.5$^{\prime}$ \\
Spatial Dispersion ($\arcsec$/pixel)			& 0.69 					& 0.40 / 0.42 				        & 0.08 \\
Spectral Resolution $R$					& 3600 							& 2500						 	    & 3744 \\
Pixel Size	($\mu$m)	                & 24 							& 13.5 							    & 15 \\
\textbf{Typical Site Seeing} ($\arcsec$)   & \textbf{1.5}                 & \textbf{1.5}                     & \textbf{0.6-1.0} \\
\hline\hline
\end{tabular} }
\caption{Instrumental setup of telescopes and detectors used for data collection. The APO ($2^{\rm nd}$) column describes both the blue and red channels for relevant parameters.}
\label{tab:telescope_info}
\end{center}
\end{table*}

In 2008, we embarked on a long-term program to acquire homogeneous long-slit spectra of VCGs on 4-m/8-m class telescopes. 
Deep long-slit spectroscopy was acquired for 138 SHIVir galaxies, including 40 VCGs by the ACS Virgo Cluster Survey \citep[ACSVCS]{Cote2004,Ferrarese2006} team
during an 11-night run in 2003 using both the 2.1-m and the Mayall 4-m telescopes at the Kitt Peak National Observatory (KPNO). 
The rest (98 VCGs) were observed by us over the period of 2008-2015 using the following instruments: the Ritchey-Chr\'{e}tien Focus Spectrograph (KPC-007 grating) on the Mayall 4.0-m telescope at the KPNO, PI: S. Courteau; the Dual Imaging Spectrograph (B1200/R1200 grating) on the ARC 3.5-m telescope at the Apache Point Observatory (APO), PIs: J. Holtzman, J. Dalcanton; and the Gemini Multi-Object Spectrograph (long-slit mode, B1200 grating) on the Gemini-South 8.2-m telescope \citep{Hook2004}, PI: N. Ouellette. 
The full technical specifications of the instruments used are listed in \Table{telescope_info}. 
The SHIVir spectra in 2008-2013 were collected on-site and remotely from the APO and KPNO observatories. 
Observations from 2014 and 2015 were collected via queue observing at the Gemini-South telescope (programs GS-2014A-Q-46 and GS-2015A-Q-79).

LTGs required $\approx 20-30$ minute integrations per galaxy on 4-m telescopes, while ETGs used $1-3$ hours per galaxy on $4-8$m telescopes. 
The exposure times were scaled as a function of visual magnitude to guarantee a minimum signal-to-noise ratio (SNR per spectral bin $> 10$) needed to extract dynamics. 
Higher levels of SNR were required for gas-poor galaxies. 
Our analysis of SNR requirements for ETGs is presented in \se{snr_req}. 

\subsection{Additional Data}
\label{sec:add_data}

The SHIVir dynamical catalogue is complemented by many literature sources: \citet{Fouque1990,Rubin1997,Rubin1999,Geha2003,Cote2004,vanZee2004,Chemin2006,Cappellari2011a,Haynes2011,Toloba2011} and \citet{Rys2014}. 
All data presented in this paper concern galaxies included in the SHIVir catalogue (designated as VCC/SDSS in \se{Catalog}). 
Available velocity dispersions and rotational velocities from these sources were used to compute the circular velocities (as described in \se{dyn_mass}). 
These are found in the ``Masses'' table found in our supplementary material, the first few rows of which are reprinted in \Table{masses}. 
Many VCGs in our catalogue have kinematic values from multiple sources. 
If multiple entries are available for a galaxy target, we use a statistical weighted average to compute a circular velocity. Details on the consistency of circular velocity measurements across different catalogue sources, including our own novel measurements, can be found in \se{dyn_mass}.

\section{Data Reduction}
\label{sec:Reduction}

Our SHIVir spectroscopic sample presented in \se{SHIVir_Spec} relies on the reduction of original spectra collected as part of our multi-year campaign. 
Intermediate-aperture (e.g., ${\sim}4$m) telescopes enabled spectroscopic observations of all possible LTG targets from the $H$-band SHIVir sample and all ETG targets brighter than $m_{b} = 13.5$.  
Fainter targets required $8-10$m class telescopes. 
Consequently, data were first collected at KPNO and APO, followed by observations with the 8-m Gemini-South telescope.  

An integral part of the SHIVir dynamical database was collected as part of the ACSVCS survey \citep{Cote2004}, and reduced by Elena Dalla Bont\`{a} (Universit\`{a} di Padova), Patrick C\^{o}t\'{e} and Laura Ferrarese (NRC Herzberg Institute), and Eric Peng (Beijing Astrophysical Centre). 
These data were integrated into the SHIVir spectroscopic subcatalogue. 
Their reduction is described in \citet{Cote2004} and \citet{Jordan2004}. 
We describe the reduction of data collected primarily for the SHIVir survey below.

\subsection{KPNO \& APO Data}

Data from the KPNO and APO telescopes were similarly formatted, and were thus reduced using the same procedures. 
The APO double spectrograph generated spectra in a blue and a red channel whereas the KPNO spectrograph used a single channel. 
The preliminary data reductions were performed using relevant \texttt{XVISTA}\footnote{\texttt{XVISTA} is maintained by J. Holtzman; see \href{http://astronomy.nmsu.edu/holtz/xvista/index.html}{this website} for documentation.} procedures. 
These included dark and bias frame subtraction, geometric and flexure correction, and wavelength calibration. 
Orthogonality of the spectrum (and for all other novel SHIVir spectra described in this section) was verified by inspecting the skyline centroids of skylines over the height of each frame.
All galaxy spectra result from at least three separate co-added integrations to enable a first round of cosmic ray removal using the median of these frames.

Remaining cosmic ray removal used the \texttt{la\_cosmic} program \citep{vanDokkum2001}. 
Cosmic-ray-free noise-variance maps were then created for the assessment of flux-calibration uncertainties.
Skylines and background sky emission were removed from each spectral frame without affecting the galaxy's spectrum.
A sky background frame was created by fitting a 1$^{\rm st}$- or 2$^{\rm nd}$-order polynomial along each column with the sky-only sections of the spectrum. 
The order function with the lowest $\chi^{2}$ was retained. The resulting sky frame was subtracted from the spectrum. 
The (non-obstrusive) galaxy continuum was not removed to keep spectral noise low. 
Finally, all frames were flux calibrated using standard stars with publicly available flux-calibrated spectra from the MILES catalogue \citep{SanchezBlazquez2006,FalconBarroso2011}.  
These standard stars were observed during each observing run. 

\subsection{Gemini Data}

Gemini data were fully reduced using the IRAF \texttt{gmos/longslit} suite of tools. 
Bias frames were taken during our observing run\footnote{All data are maintained at the \href{https://archive.gemini.edu}{Gemini Observatory Archive}.}, and bias subtraction and flat-fielding were performed on both sets of science (galaxy and standard star) frames. 
A wavelength solution was built using CuAr (copper and argon) arcframes whose linelists for accurate line matching were available in the standard IRAF distribution. 
The wavelength solution was applied to all science frames and sky subtraction was performed in a manner analogous to the KPNO and APO spectra.

Each final Gemini galaxy spectrum results from the combination of at least three integrations, much like our APO and KPNO observations. 
This enabled a first round of cosmic ray removal, and the creation of an initial noise-variance map to determine flux calibration uncertainties. 
Furthermore, because the Gemini longslit has 2 gaps where signal is lost, the different frames for a given galaxy were all dithered in the radial direction to achieve maximal spatial coverage. 
The frames were co-added to produce the final reduced spectrum used in our data analysis. 
Subtracted sky maps were added to our noise-variance maps, as for APO and KPNO data, to produce final variance maps from which uncertainties were inferred.


\section{Kinematics Extraction}
\label{sec:DataAnalysis}

Our spectroscopic data have been extensively tested to produce a most reliable and comprehensive dynamical database for Virgo cluster galaxies. 
We now review the methodology used in determining kinematics first described in \S 2 of O17 --- rotational velocity for LTGs and velocity dispersion for ETGs --- and dynamical masses. 
These and the photometric parameters described in \se{Photometry} are then combined to construct Virgo cluster scaling relations (O17).

\subsection{Rotational Velocity}
\label{sec:rotvel}

\begin{figure}
\centering
\includegraphics[width=0.45\textwidth]{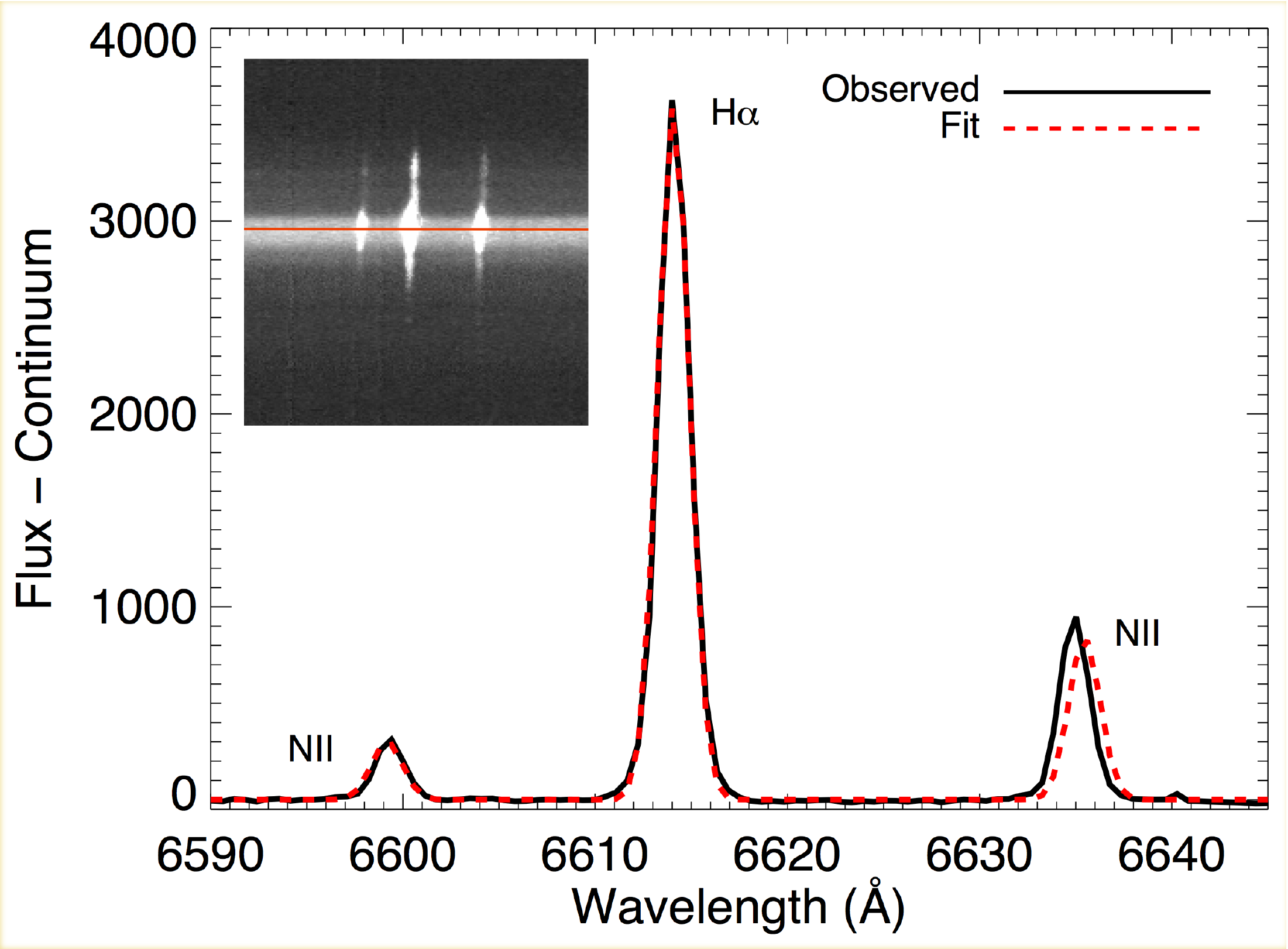}
\caption{Triple Gaussian fit of the \ntwo-\ha-\ntwo\ emission complex at the centre of VCC 692. 
The observed SED and fit are shown as a black and red dashed line, respectively. 
The inset is the observed 2D galaxy spectrum collected at APO; the red line indicates the extracted spectral row.}
\label{fig:tripgauss_rc}
\end{figure}

We extract the SHIVir rotational velocities by fitting a triple Gaussian function over the \ntwo-\ha-\ntwo\ emission complex for each spectral row (corresponding to a specific galactocentric radius). 
\Fig{tripgauss_rc} shows the triple Gaussian fit for the centre of galaxy VCC 692.
As previously mentioned, the continuum was not removed during data reduction since it is not obstrusive.
The continuum was removed in \Fig{tripgauss_rc} only for clarity. 
In this case, the small mismatch between the observed Gaussian peaks and the fitted peaks is due to wavelength calibration uncertainties.
Emission line peaks and uncertainties are computed using an intensity-weighted centroid method \citep{Courteau1997} and converted from wavelength to velocity space using the Doppler equation:

\begin{equation}
 \frac{V^{\rm i} - V_{0}}{c} = \frac{\lambda^{\rm i}_{\rm obs} - \lambda_{\rm em}}{\lambda_{\rm em}},
\end{equation}
 
\noindent where $c$ is the speed of light, to find the row's rotational velocity $V^{\rm i}$ relative to the galaxy's bulk motion $V_{0}$. 

The rest-frame \ha\ emission line centred at $\lambda_{\rm em}=6563$~\AA\ is thus compared with the observed centroid for the i$^{\rm th}$ row $\lambda^{\rm i}_{\rm obs}$. 
The central wavelengths for the \ntwo\ lines are taken to be 6548 and 6584~\AA.

A preliminary estimate of the location of the true dynamical centre of the galaxy is challenging since the emission line peak may not match that true centre \citep{Franx1991,Andersen2001,Fridman2005}. 
To find the kinematical galaxy centre, we first locate the spatial peak in pixel value along the emission line. 
This spatial point is set as a temporary galaxian centre. 
While this is often a decent estimate of the galaxy's centre, the brightest point in the emission line is not necessarily indicative of the true galaxy potential well. 
Moreover, some of our galaxies do not show a defined emission peak, especially for those galaxies exhibiting an \ha-\ntwo\ reversal \citep{Rubin1986,StorchiBergmann1991,Courteau1997}, where we find the \ha\ emission to be fainter than the \ntwo\ emission at the galaxy centre. 
As a remedy, RCs are systematically folded about different location along their curve. 
To find the velocity centre, symmetry is assumed between both sides of the galaxy. 
The $\chi^{2}$ of the fit for both sides of the galaxy folded together is computed for each of these folding points. 
The folding point resulting in the smallest $\chi^{2}$ value is chosen as the centre of the galaxy. 
The folding point is typically found to be $0.5-5$\arcsec away from the photometrically-determined centre.

The extracted extended RCs were fit using a multi-parameter function (\citealp[]{Courteau1997}):

\begin{equation}
\label{eq:mpf}
V(R) = V_{\rm crit} \ \frac{1}{(1 + x^{\gamma})^{1/\gamma}},
\end{equation}

\noindent where $x = R_{\rm t}/R$, $\gamma$ controls the degree of sharpness of the RC's turnover, $V_{\rm crit}$ is the asymptotic maximum velocity, and $R_{\rm t}$ is the radius at which the transition between the rising and flat parts of the RC occurs\footnote{Other similar fitting functions exist; see e.g., \citet{Bertola1991,Verheijen2001,Giovanelli2002,Noordermeer2007}.}. 
Rotational velocities used in scaling relations such as the Tully-Fisher relation (TFR: \citealp{Tully1977}, \citealp{Courteau2007b}, O17, \citealp{Stone2021a}) can be measured at different locations along the fitted RC model. 
The following measured rotational velocities are listed in the ``Kinematics'' table found in the supplementary material, the first few rows of which are shown in \Table{kinematics}: 
$V^{\rm c}_{2.2}$ measured at 2.15 disc scale lengths which corresponds to the peak rotational velocity of a pure exponential disc \citep{Freeman1970,Courteau1997}, 
$V^{\rm c}_{23.5}$ measured at the isophotal radius $R_{23.5}$, 
corresponding to the $i$-band 23.5 mag arcsec$^{-2}$ isophotal level, 
and $V^{\rm c}_{\rm max}$ measured along a model fit at the last radial point where \ha\ emission is still detected. 

Virgo RCs appear to be truncated, likely due to the Virgo cluster environment (see Fig. 1 of O17). This feature is detected in all kinematic studies of Virgo RCs in \ha\ and \ion{H}{I} \citep{Crowl2007,Koopmann2004,Chung2009,Cortese2010}. 

The rotational velocities are all corrected for inclination using SHIVir photometric isophotes from the reddest, and thus least dust-obscured, images. 
This correction is indicated by the superscript ``c'' on all LTG kinematic parameters. 
The RMS difference between our reduced SHIVir rotational velocities and those taken from the literature for overlapping galaxies is ${\sim}10-15\%$, with no noticeable systematic bias.

\subsection{Velocity Dispersions}
\label{sec:vd_extr}

\begin{figure}
\centering
\includegraphics[width=0.45\textwidth]{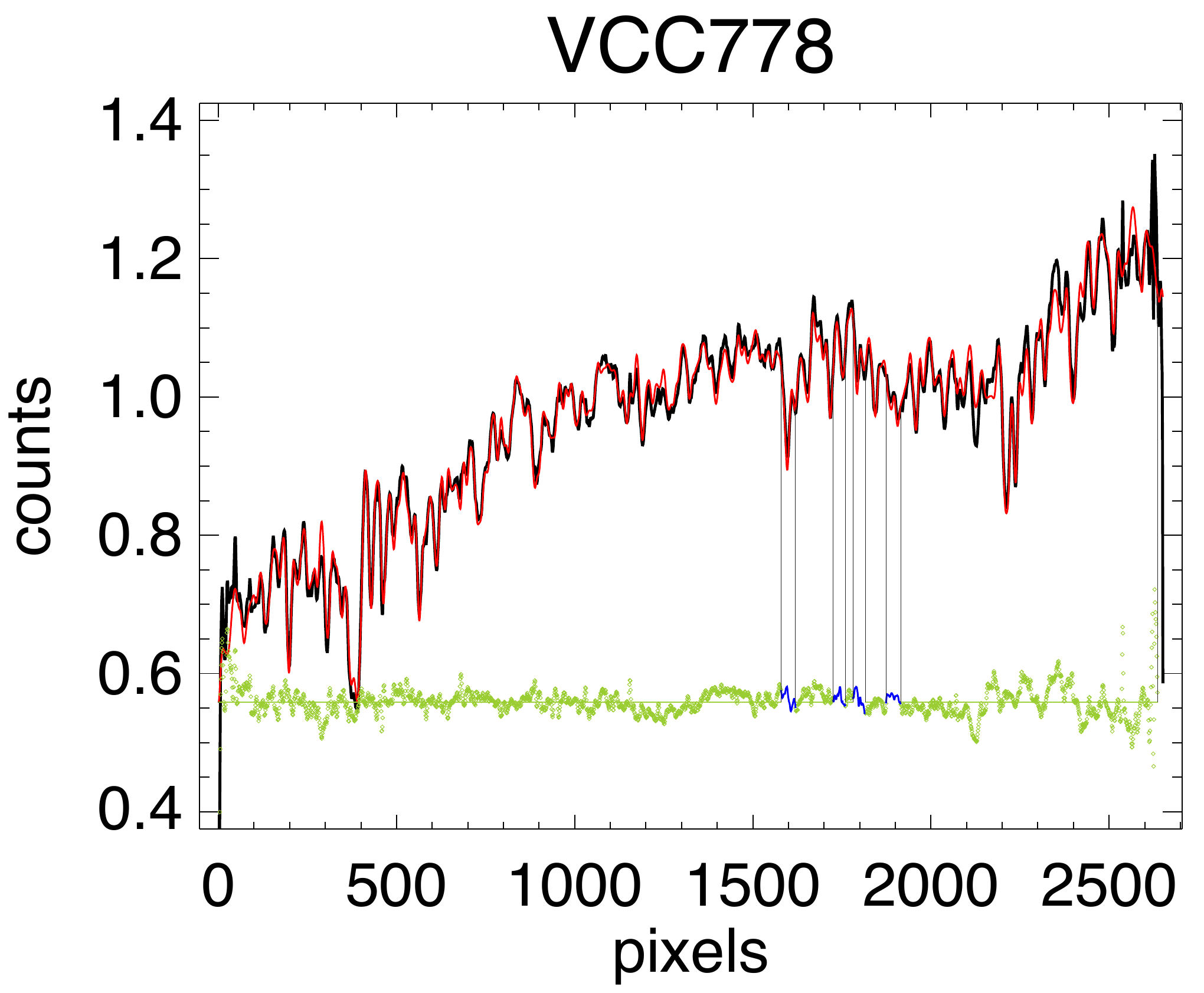}
\caption{SED of galaxy VCC 778 in pixel space, with \texttt{pPXF}'s best-fit combination of stellar templates convolved with a Gauss-Hermite function overplotted in red. Residuals (normalised counts in arbitrary units) between the observed spectrum and the fit are shown in green (in blue where the spectrum was masked due to emission lines).}
\label{fig:vcc778_spec}
\end{figure}

The velocity dispersion, $\sigma$, is the nominal kinematic parameter for pressure-supported, gas-poor systems such as ETGs. 
Stellar kinematics were extracted from ETGs' absorption spectral features using a penalised pixel-fitting method, \texttt{pPXF} \citep{Cappellari2004,Cappellari2017}. 
This algorithm constructs a best-fit linear combination of stellar templates to the observed galaxy spectrum and computes the first four moments of a Gauss-Hermite function needed to convolve this combination for an accurate match. 
The 2$^{\rm nd}$ moment corresponds to the system's stellar line-of-sight velocity dispersion (LOSVD). 
We used stellar templates from the MILES library \citep{SanchezBlazquez2006,FalconBarroso2011a} to extract integrated and resolved velocity dispersions. 
\Fig{vcc778_spec} shows a sample spectrum of galaxy VCC 778, with \texttt{pPXF}'s best fit overplotted. More details about \texttt{pPXF} and any associated biases due to certain input parameters are given in \se{ppxf_syst}. 
For integrated measurements, spectra collapsed spatially over a range of slit lengths (e.g. $R_{\rm e}/4$, $R_{\rm e}$, $2 R_{\rm e}$, etc.) were also fit by \texttt{pPXF}. 
These enabled the measurement of a dynamical mass enclosed within each aperture (see \se{dyn_mass}). 
They were also used to characterise the dependence of the Fundamental Plane (FP)'s scatter on VD aperture size in O17. 

The wide range of aperture shapes (circular, elliptical, slit, etc.) used in kinematic studies and radially-changing VD profiles complicates the uniform comparison of measured and published velocity dispersions (VDs).  
Aperture corrections to account for the different VD profile shapes were first applied by \citet{Jorgensen1995} in the context of FP studies;
these authors corrected VDs to a circular aperture of diameter equal to 1.19 h$^{-1}$ kpc, or 3.4$\arcsec$ at the distance of the Coma cluster (99 Mpc). 
They also computed VDs within a circular aperture of diameter $R_{\rm e}/4$.  
Their methodology has been applied by others to VCGs \citep{Gavazzi2003} and other galaxies. 
A physical aperture (in kpc) will clearly sample different parts of a galaxy depending on its shape. 
Conversely, an aperture based on projected angles, such as arcseconds, will also sample different parts of a galaxy depending on its distance. 
A scalable metric such as one based on $R_{\rm e}$ might favourably alleviate such sampling effects \citep{Cappellari2006}. 
In cases where this was applied, an aperture correction was needed since their spectra often did not extend out to $R_{\rm e}$. 
However, these corrections can still be fraught with uncertainty and usually rely on empirical fits of datasets such as assuming that the relation between VD and galactocentric radius follows a power-law. 
We chose not to correct our VDs for any assumed VD profile or aperture bias. 
Our study benefits from VDs measured within larger apertures than found in the current literature, and the need for an aperture correction is thus lessened. 

For resolved velocity dispersion measurements, from which dispersion profiles are built (see \se{kinprofiles}), spectral rows are binned radially over 3 pixels for each radial data point to allow for sufficient dispersion to be observed. 
In the dimmer outskirts of a galaxy, rows are binned until the SNR level of 50/\AA\ is reached; this criterion is revisited in \se{snr_req}. 
The dispersion measurements characterise the galaxy kinematics locally. 

An integrated central velocity dispersion, $\sigma_{0}$ (taken within an aperture of $R_{\rm e}/8$), was measured for 131 VCGs, whereas an integrated effective velocity dispersion $\sigma_{\rm e}$ was measured for 128 VCGs (see Fig. 3 of O17), 88 of which are ETGs. The integrated velocity dispersion measurements measured at the centre, $\sigma_{0}$, within $R_{\rm e}/4$, $\sigma_{R_{\rm e}/4}$, within $R_{\rm e}$, $\sigma_{R_{\rm e}}$, and within $2 R_{\rm e}$, $\sigma_{2R_{\rm e}}$, are presented in our ``Kinematics'' table found in the supplementary material, the first few rows of which are shown in \Table{kinematics}.

\subsubsection{Velocity Dispersion Systematics}
\label{sec:ppxf_syst}

While \texttt{pPXF} is a popular algorithm for extracting galaxies' stellar kinematics, and has been used in a number of IFS surveys (e.g. \citet{Emsellem2004} for SAURON, \citet{Cappellari2011a} for \atlas, \citet{FalconBarroso2017} for CALIFA, \citet{vandeSande2017} for SAMI, \citet{Westfall2019} for MaNGA), final kinematic results depend acutely on the range of adopted configuration parameters.  
In this section, we present a systematic analysis of such parameters: SNR, stellar templates, and masked emission lines.

\paragraph{Signal-to-Noise Requirements}
\label{sec:snr_req}

We wish to calibrate the dependence of stellar kinematics measured with \texttt{pPXF} on the SNR of our observed spectra.  
This will justify the minimum SNR level of 50/\AA\ in cases where radial binning is required in order to extract a VD profile in the outskirts of galaxies.

To study the SNR dependence of our extracted kinematics, we created mock galaxies by adding spectra of stars typically found in ETGs and artificially broadened the stacked spectra to an imposed dispersion value for recovery by \texttt{pPXF}. 
We used our mock spectra to determine the level of noise beyond which a VD cannot be reliably recovered. 
To simulate ETGs, we co-added the spectra of 1 G-type and 3 K-type stars from the MILES library \citep{SanchezBlazquez2006,FalconBarroso2011a}, and convolved the resulting stack with a Gaussian function of width: 100, 150, 200 and 250 km~s$^{-1}$.

\begin{figure}
\begin{center}
    \includegraphics[width =0.45\textwidth]{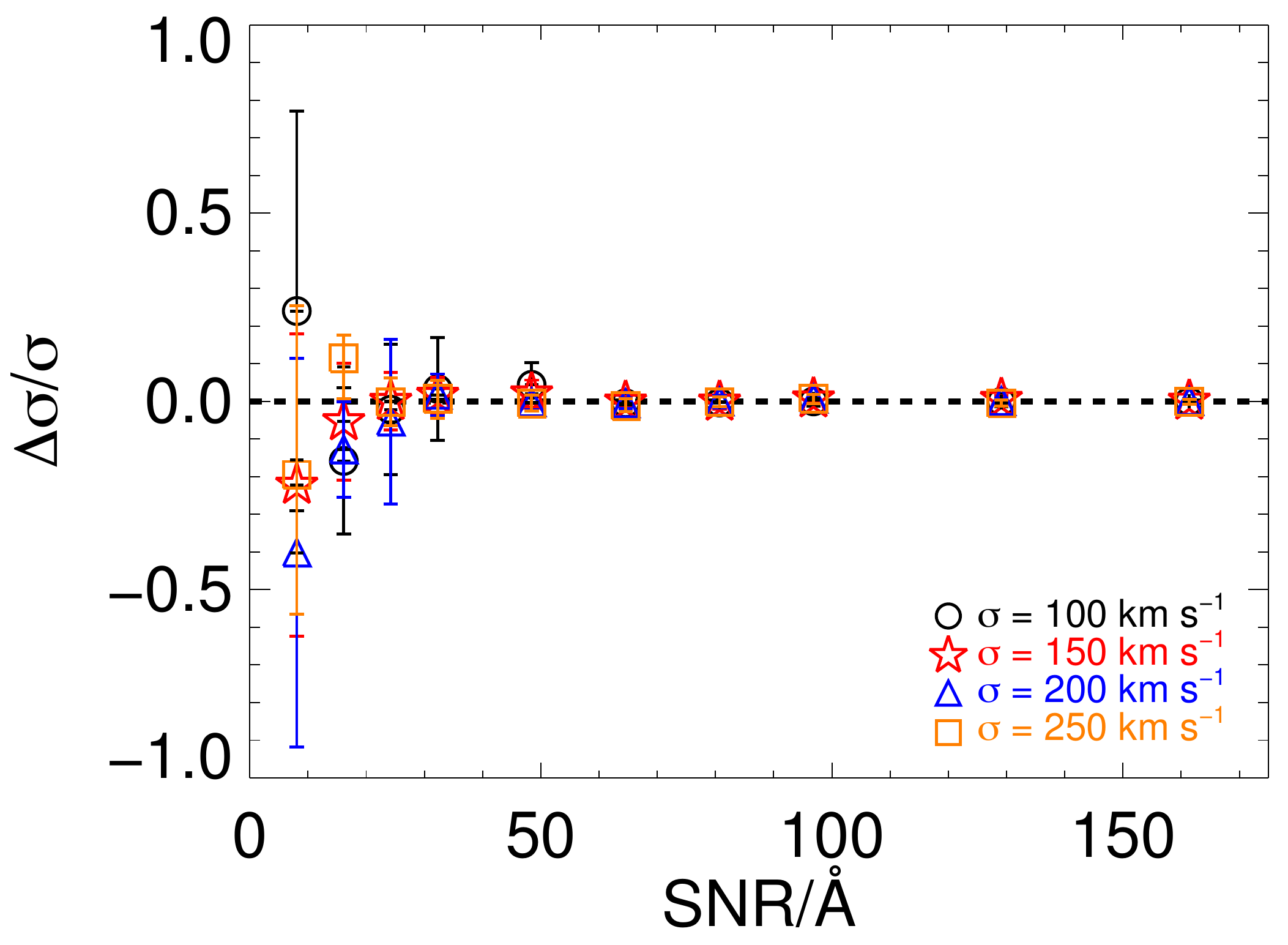}
    \caption{Relative difference between true and extracted velocity dispersion of simulated galaxies, $\Delta\sigma/\sigma$, as a function of SNR. 
    The 1-$\sigma$ level uncertainties are plotted for each point. Four different levels of broadening (100, 150, 200 and 250 km~s$^{-1}$) were tested.}
\label{fig:snr_deltasig}
  \end{center}
  \end{figure}
  
Since the absolute flux level of the input SED is inconsequential in the \texttt{pPXF} algorithm, the number of stars (i.e. relative intensity of our spectra) is also irrelevant. 
In fact, 1D spectra are often divided by their own median to avoid possible numerical issues and for proper normalisation. 
We then degraded our composite mock spectra with varying levels of noise (to chosen SNR levels), and fed them into \texttt{pPXF}. 
The recovered VD should be close to the width of the convolution Gaussian. 
Noise was added to the composite spectrum according to the standard definition of SNR:

\begin{equation}
\sigma_{noise} = \frac{Signal}{SNR}.
\end{equation}

\noindent The noise added to the spectrum at each wavelength bin was sampled randomly from a Gaussian distribution centred at 0 with a spread equal to the aforementioned $\sigma_{\rm noise}$. 

The relative offsets between the \texttt{pPXF} results and the expected VDs as a function of SNR are plotted in \Fig{snr_deltasig}. 
All four tested velocity dispersions are shown. 
Results are fairly consistent in all four cases and match those of \citet{Toloba2011}. 
Error bars are negligible at SNR levels of 65/\AA\ and above. 
A less stringent cutoff at an SNR level of 30/\AA\ yields VD errors of 15\%.  
Below this threshold, VD errors diverge.

In this work, we have applied an SNR cutoff of 50/\AA\ to limit VD errors to 5\%. 
A higher cutoff would yield more reliable kinematic measurements, but it would also result in sparse or prematurely truncated VD profiles and thus hinder our extended kinematic profile analysis in \se{kinprofiles}.

\paragraph{Stellar Template Choice}
\label{sec:temp_dep}

\pPXF requires approximately 100 to 200 stellar templates, in anticipation that $\sim$10\% of them will be retained in the final galaxy spectrum fit.  
Not only does the choice of stellar templates used in \pPXF greatly affect the final results, it also hampers the analysis efficiency.
A fully optimised stellar template database should yield most accurate results in the least amount of processing time.

To test for template catalogue bias, we attempted to recover kinematics from the same simulated galaxies built for our SNR analysis using three different stellar template catalogues. 
We included an additional broadening level of $50 \kms$ to consider the effects of template selection on dwarf galaxies. 
First, we selected 20 random stellar spectra from the MILES catalogue, only recording their spectral types and ensuring that they are not uncommon or unimportant in ETGs (i.e. no variable stars, O-type stars, etc.). 
We called this the ``Random'' set.  
Second, we used 50 stellar spectra hand-picked from the MILES catalogue by colleagues Lorenzo Morelli and Enrico Maria Corsini (priv. comm. Universit\`{a} degli studi di Padova) to build the ``Padova'' template set.  
Spectral types and MILES ID were not directly available for these templates.  
A least-squares spectral matching code, to find the best match for each of these spectra with the full MILES catalog, shows that the Padova set is an even mixture of F- and K-type stars. 
Third, and finally, we used the full MILES library (985 stellar templates) as our final set: we called this the ``MILES'' set.

 \begin{figure}
    \begin{center}
    \includegraphics[width =0.45\textwidth]{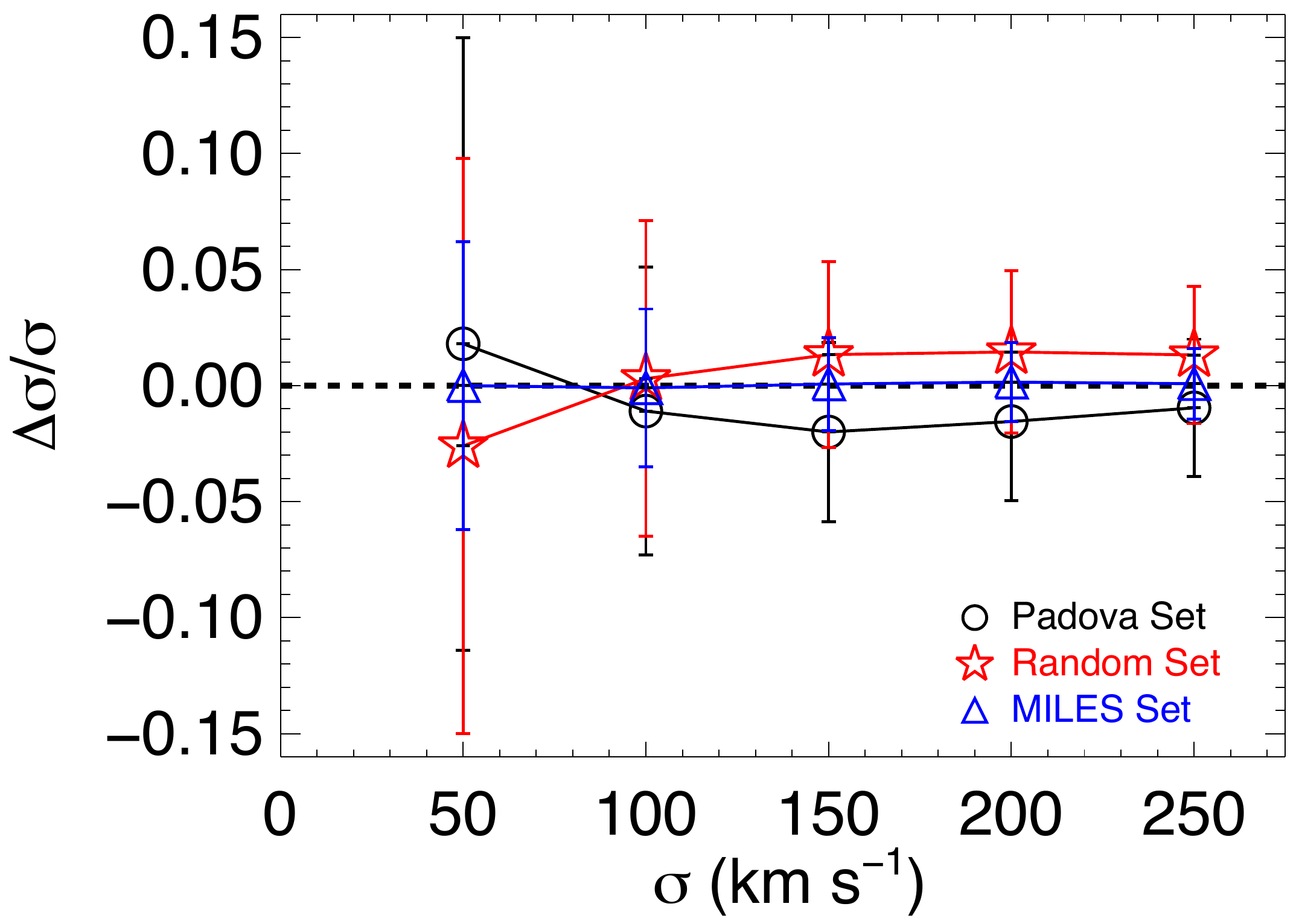}
    \caption{Relative difference between true and extracted velocity dispersion of simulated galaxies, $\Delta\sigma/\sigma$, as a function of the set of stellar templates used. The 1-$\sigma$ level uncertainties are plotted for each point. Five different levels of broadening (50, 100, 150, 200 and 250 \kms) were tested.}
\label{fig:temp_deltasig}
   \end{center}
\end{figure}

Given our chosen SNR minimum level, we tested galaxies with a simulated SNR of 50/\AA. 
Similarly to \Fig{snr_deltasig}, the relative differences between true and extracted VD for five mock galaxy sizes for all three template sets are shown in \Fig{temp_deltasig}. 
We found the uncertainties to be noticeably smaller for the ``MILES'' set, and the uncertainties for the ``Padova'' and ``Random'' sets to be comparable to each other. 
The ``MILES'' set also seemed to produce the most accurate kinematics for the entire mass range. 
In cases where literature VDs were available for comparisons, we also generally found a better match between these published results and our results extracted by \pPXF using the ``MILES'' set.  
Providing \pPXF with an extensive template catalogue allowed for the best spectral fits. 

Ideally, we should fit galaxy spectra with as comprehensive a stellar template library as possible to produce accurate kinematics.  
However, the size of the MILES library and the large number of radial LOSVD measurements required to build a resolved VD profile would make this effort computationally prohibitive. 
In order to streamline our operations, we used a spatially collapsed (``radially mashed'') spectrum over an aperture of $2 R_{\rm e}$ for each galaxy for which \texttt{pPXF} selects an optimal stellar template sub-catalogue from the entire MILES library. 
Barring any notable radial gradients of the stellar populations (of which we found none), this yielded a sub-catalogue of approximately $15-20$ stellar templates chosen from the MILES library for each galaxy that produces stable fits in a time-efficient manner. 
Each galaxy thus has a reduced {\it customised} (rather than random) library of templates with which we can perform spectral fitting. 
This stellar template configuration is a good compromise between efficiency and accuracy and was used throughout this work. We compared the VD results based on this approach with those retrieved using the full MILES library for a few representative galaxies in the SpecSHIVir catalogue and found them to match within a few percent.

\paragraph{Masked Emission Regions}

\begin{figure}
\begin{center}
    \includegraphics[width =0.45\textwidth]{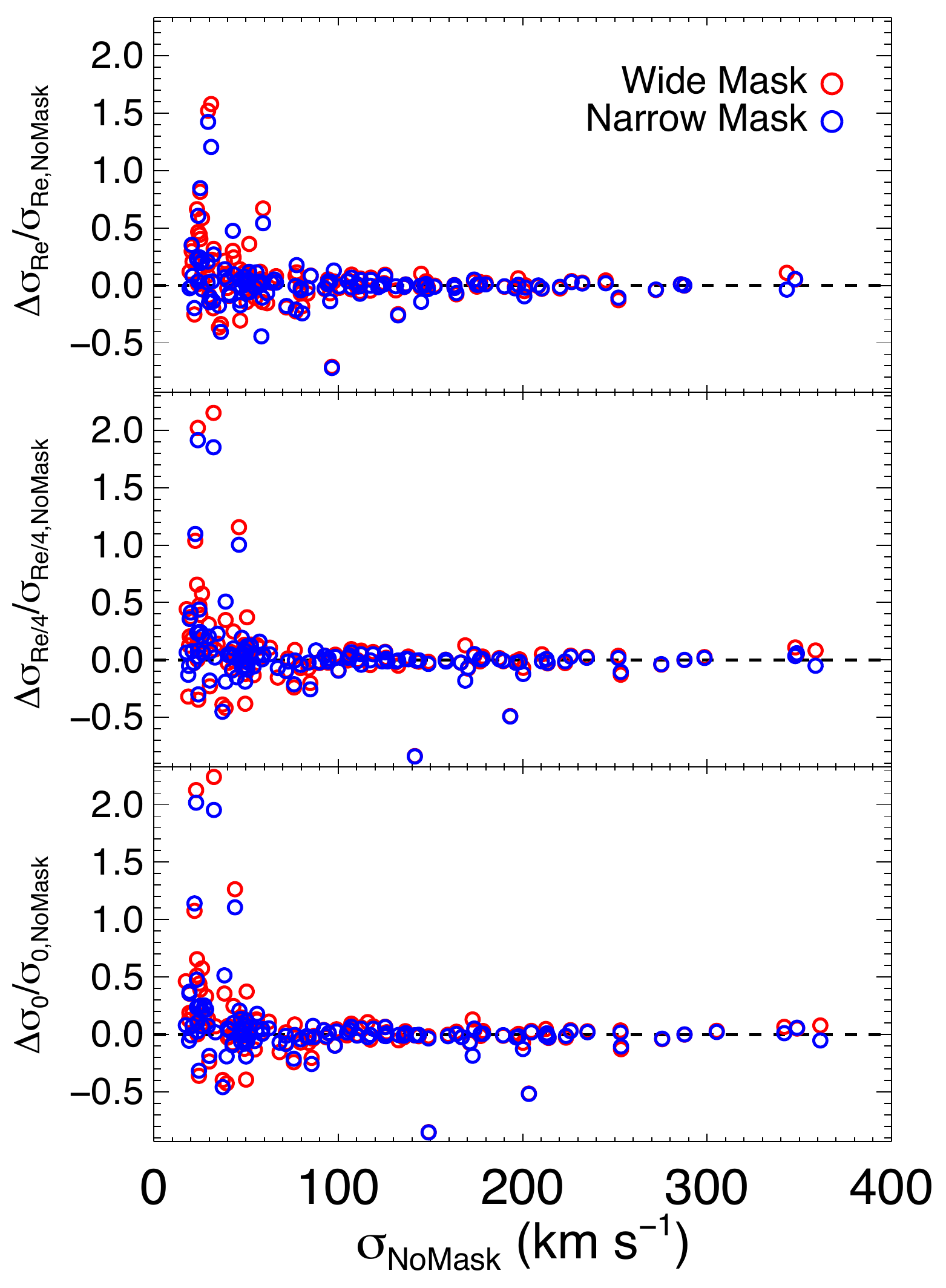}
    \caption{Comparison of extracted VDs using no mask, narrow masks, or wide masks in \texttt{ppXF}. Central VD $\sigma_{0}$ (bottom), VD measured inside $R_{\rm e}/4$, $\sigma_{R_{\rm e}/4}$ (middle), and VD measured inside $R_{\rm e}$, $\sigma_{R_{\rm e}}$ (top), are shown. 
    Comparison between ``No Mask'' and ``Wide Mask'' configurations are shown in red, comparison between ``No Mask'' and ``Narrow Mask'' configurations are shown in blue. 
    The black dashed line indicates zero residuals.}
\label{fig:mask_comp}
  \end{center}
\end{figure}

The spectral fitting algorithm used in \texttt{pPXF} relies on stellar absorption lines to retrieve kinematics which can be contaminated by varying amounts of gas emission found in many galaxies. 
Therefore, the observed galaxy spectrum analysed by \texttt{pPXF} must be partially masked for proper fitting.  
Various emission lines were masked to enable reliable \texttt{pPXF} fits (depending on the wavelength range of each spectrum): H$\beta$ (4862.68~\AA), the \othree\ triplet (4932.603~\AA, 4960.295~\AA, 5008.24~\AA), \hetwo\ (5412.00~\AA), \ntwo\ (5755.00~\AA), \heone\ (5876.00~\AA), and the \oone\ doublet (6302.05~\AA, 6365.54~\AA). Additionally, a width of the mask applied to each emission line must be determined.

Finding an optimal width for each line that works for all our VCGs is challenging; for full masking, the width of the mask must be adjusted for the VD of the galaxy which affects the intrinsic broadening of its emission lines. 
To verify whether the mask selection systematically biased our results, three different mask configurations were tested: no masks, narrow masks, and broad masks. 
Narrow masks were defined by strictly masking the emission lines of most galaxies.
The half-width of these masks chosen for each line range between 500 and 1500 \kms, or 8 and 30~\AA.
For more emissive or turbulent galaxies, this proved insufficient in order to match the broad lines or noise between neighbouring lines. 
Broad masks were allowed to overlap in order to fully cover the entire region between 4830 and 5030~\AA.
Lines beyond the \othree\ triplet had masks with half-widths ranging between 1200 and 2000 \kms, or between 22 and 40~\AA.
A broad mask covering the region between 5400 and 6400~\AA\ was used for our most massive galaxies ($\sigma_{\rm e} \sim 350$ \kms).
The three mask configurations were used on all the SHIVir galaxies (for which we have available raw spectra). 

VDs extracted with narrow and broad masks were compared to VDs extracted without masks in \Fig{mask_comp}.
The comparison used VDs measured at the centre, $\sigma_{0}$, and within apertures of $R_{\rm e}/4$, $\sigma_{R_{\rm e}/4}$, and $R_{\rm e}$, $\sigma_{R_{\rm e}}$. 
We have defined $\Delta \sigma = \sigma_{\rm Wide} - \sigma_{\rm NoMask}$ or $\Delta \sigma = \sigma_{\rm Narrow} - \sigma_{\rm NoMask}$, depending on the colour of the datapoint plotted. 
For most galaxies with $\sigma < 100$ \kms, our results depend greatly on whether emission line regions were masked or not. 
This is likely explained by the smaller broadening of absorption lines relative to the larger dispersion from emissive gas in smaller ETGs and early-type spirals in this velocity regime. 
The discrepancy appeared to be smaller when VDs were measured within larger apertures, implying that the disruptive emission is stronger at or near the galaxy centre, and can be mitigated by considering the galaxy's outskirts. 
The lack of masked regions generally led to underestimated VD measurements, compared to both narrow and wide masks.

\begin{figure}
\begin{center}
    \includegraphics[width =0.45\textwidth]{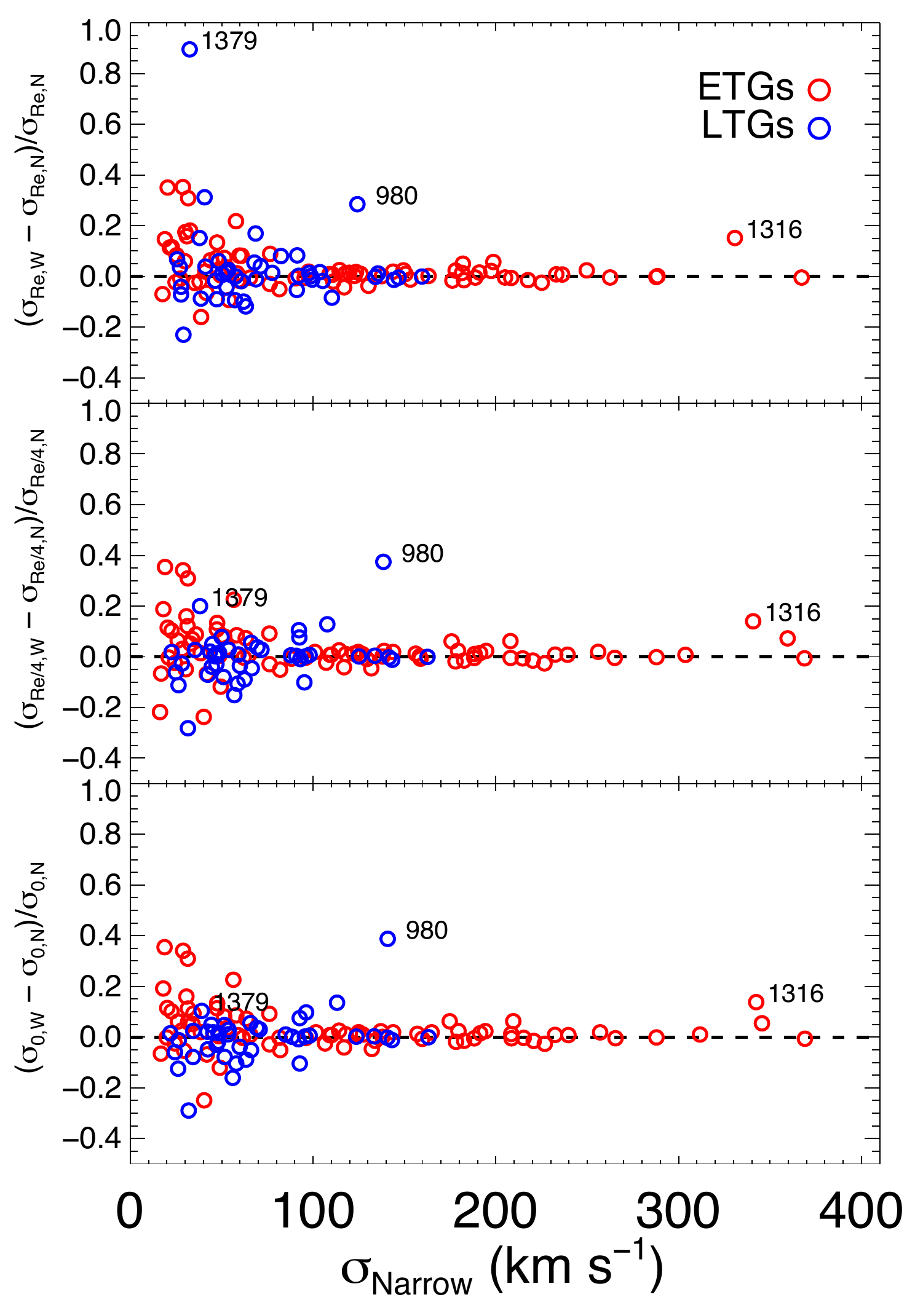}
   \caption{As in \Fig{mask_comp}, but comparing wide and narrow masks only.
   ETGs and LTGs are shown in red and blue, respectively. 
   Outlying galaxies' VCC numbers are marked in black.}
\label{fig:mask_comp2}
  \end{center}
  \end{figure}
  
\begin{figure*}
\begin{center}
    \includegraphics[width =0.99\textwidth]{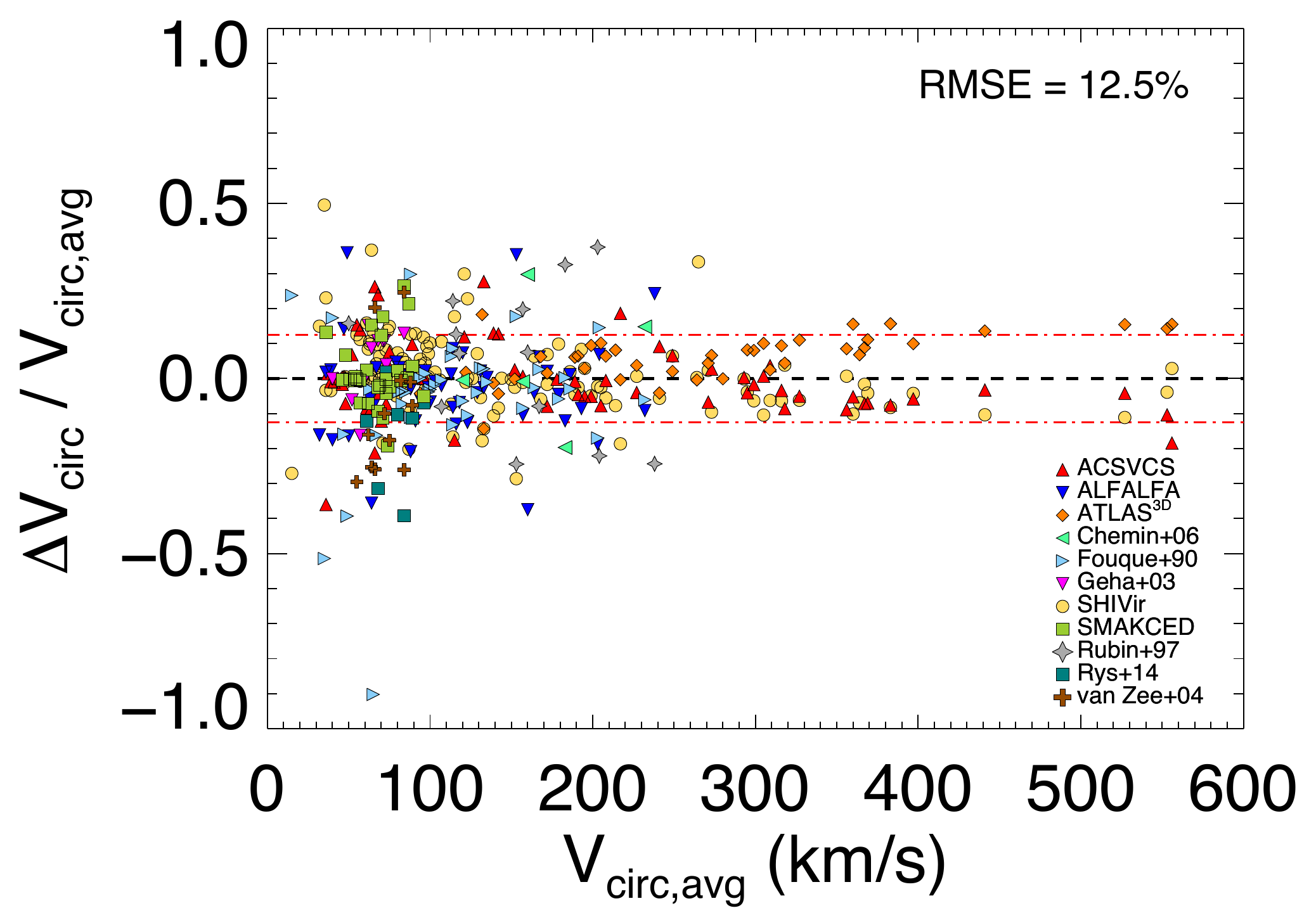}
   \caption{Relative circular velocity scatter, 
   $\Delta V_{\rm circ}/V_{\rm circ, avg} = (V_{\rm circ,avg} - V_{\rm circ,catalogue})/V_{\rm circ,avg}$, 
   as a function of the weighted average $V_{\rm circ,avg}$ calculated for all available catalogue sources as listed in \Table{masses} 
   and shown in the legend (see \se{add_data} for references).
   The root mean square error (RSME), $\Delta V_{\rm circ}$ / $V_{\rm circ, avg} = $ 12.5\%, is delineated by red dash-dotted lines.}
\label{fig:vcirc_comp}
  \end{center}
  \end{figure*}
  
\begin{figure*}
\centering
\includegraphics[width=0.99\textwidth]{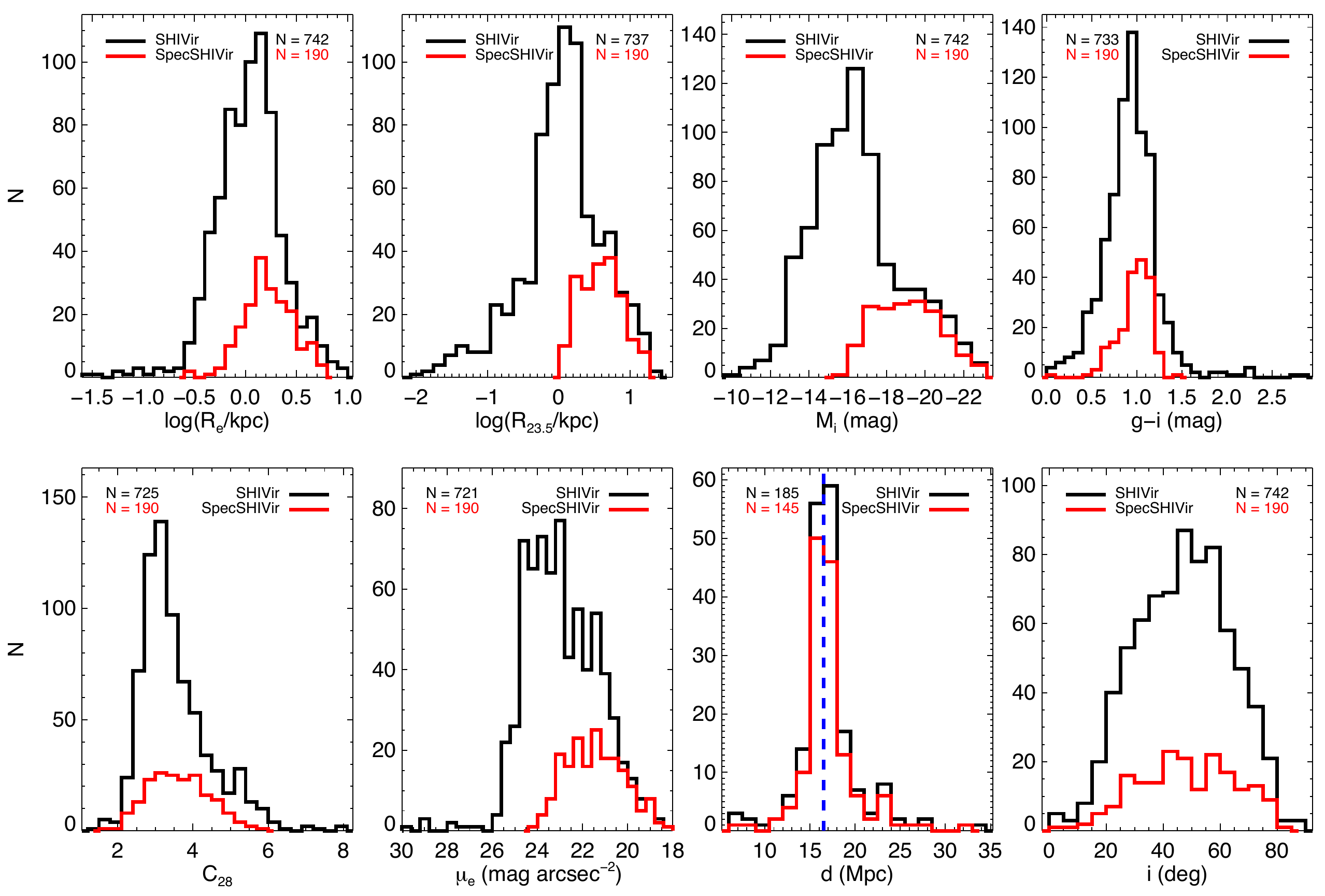}
\caption{Histograms of the full SHIVir catalogue's (black) and the spectroscopic sub-catalogue SpecSHIVir's (red) photometric parameters: effective radius, $R_{\rm e}$; 23.5 \magarc level isophotal radius, $R_{23.5}$; $i$-band absolute magnitude, $M_{i}$; $g-i$ colour; concentration, $C_{28}$; effective $i$-band surface brightness, $\mu_{\rm{i,e}}$; distance, $d$; and inclination, $i$. 
Sample sizes $N$ are indicated. 
The distance to Virgo, 16.5 Mpc, is indicated in the $d$ histogram by a blue dashed line. }
\label{fig:Photo_Hist}
\end{figure*}
  
 \Fig{mask_comp2} shows the importance of mask selection in \pPXF by comparing extracted VDs depending on the width of the imposed masked regions. 
 The scatter in VD results due to a narrow vs a wide mask is considerably smaller than the scatter shown in \Fig{mask_comp} which is based on whether there is a mask at all or not. 
 Save one exception, the uncertainty on VD between mask choices is below 15\% for galaxies with $\sigma > 100$ \kms. 
 Three outlying galaxies stand out: VCC 980, VCC 1379, and VCC 1316. 
 The first is a weakly-rotating disorganised spiral, the second is a weakly-barred spiral. 
 A rotation curve was extracted for both of these systems, and neither are included in our FP analysis in O17 due to their significant VD uncertainties. 
 The third is Virgo's massive central elliptical, M87, hosting an AGN \citep{Lea1982}. 
 Our catalogue selection generally avoided galaxies with active centres, but we observed M87 for its legacy value. 
 A visual inspection of the \pPXF fitting of M87's spectrum shows a better fit when the wider masks are used due to the system's high velocity dispersion and turbulent central region. 
 We conclude that the kinematics of smaller galaxies and galaxies with known emission regions depend sensitively on the choice and width of a mask in \texttt{pPXF}. 
 For these systems, the masks were best configured on a case-by-case basis to achieve the most accurate VD results by visually inspecting the width and strength of their emission lines. 
 For better behaved systems as determined by their placement in  \Fig{mask_comp2}, narrow masks can be applied.
 
\subsection{Circular Velocities \& Dynamical Masses}
\label{sec:dyn_mass}

Kinematics are extracted separately for LTGs and ETGs, and must be merged for a comprehensive study of all VCGs. 
Towards this end, we inferred circular velocities and dynamical masses for all 190 SHIVir VCGs.  
These included galaxies observed by us and others with kinematics published in the literature, as described in \se{add_data}.  
The procedure to compute these values is described in \S 2.4 of O17.  
It is reviewed briefly below.

Dynamical masses inside a fiducial projected radius, $R$, were inferred for rotating discs using:

\begin{equation}
\label{eq:mdyn}
M_{\rm dyn}(R) = \frac{V_{\rm circ}^{2} R}{G},
\end{equation}

\noindent where $G$ is the gravitational constant, the circular velocity is $V_{\rm circ} = V^{\rm c}_{\rm rot}=V^{\rm c}_{\rm 23.5} = V_{23.5}/{\rm sin\ } i$ for LTGs, $i$ is the inclination of the galaxy disc and the superscript ``c'' indicates that the velocity is inclination-corrected (no correction for redshift broadening is necessary for these nearby systems). 
For ETGs, analogous values of $M_{\rm dyn}$ measured inside the radius of a sphere enclosing half of the total stellar mass were computed using

\begin{equation}
\label{eq:mdyn_etgs}
M_{\rm dyn}(r_{1/2}) = c \frac{r_{1/2} \sigma^{2}_{\rm e}}{G},
\end{equation}

\noindent where the structural constant $c$ is computed from the function $c = -0.300n + 4.153$, built from the values in Table II of \citet{Courteau2014} and the S\'{e}rsic index $n$ was measured from our $i$-band bulge-disc decompositions described in \citet{McDonald2011}.
The circular velocity for ETGs was assumed to be $V_{\rm circ} = \sqrt{c}\sigma_{\rm e}$.
Dynamical masses were all computed within physical spherical radii such as the half-light radius $r_{1/2}$. 
For LTGs, we assumed that cylindrical, $R$, and spherical, $r$, radii are comparable, and therefore $r_{1/2} = R_{1/2}$. 
For ETGs, we reasoned that $r \approx (4/3) * R$ for pure stellar systems \citep{Hernquist1990,Ciotti1991}. 
If available, we also used \hi gas masses taken from the ALFALFA $\alpha.100$ catalogue \citep{Haynes2011}.
Finally, the dark matter mass $M_{\rm DM}$ can be inferred as:

\begin{equation}
M_{\rm DM} = M_{\rm dyn} - M_{*} - M_{\rm gas},
\end{equation}

\noindent where $M_{*}$ is the stellar mass computed as in \se{StellarMasses} and $M_{\rm gas}$ is the \hi gas mass.  
If $M_{\rm dyn} < M_{*}$, we set $M_{\rm DM} = 0$. 
These masses, along with our computed circular velocities $V_{\rm circ}$, are included in the ``Masses'' table in the supplementary material, the first few rows of which are listed in \Table{masses}.
This table concerns the extended SHIVir kinematic sample (SpecSHIVir, 190 galaxies) which includes VCGs with literature kinematic measurements (\se{add_data}).

As stated in \se{add_data}, we used a statistical weighted average to compute circular velocities if velocity measurements from multiple different sources were available for a given galaxy.
We have verified the consistency of measurements from different catalogues by computing the root mean square error (RMSE) of the scatter of $\Delta V_{\rm circ}/V_{\rm circ, avg} = (V_{\rm circ,avg} - V_{\rm circ,catalogue})/V_{\rm circ,avg}$, as seen in \Fig{vcirc_comp}.
The scatter decreases as $V_{\rm circ,avg}$ increases, as expected given the difficulty of extracting kinematics from lower mass galaxies.
We confirm that this is not due to a constant absolute scatter across all mass bins, and that absolute scatter actually increases slightly with mass.
The RMSE of the $\Delta V_{\rm circ}/V_{\rm circ, avg}$ scatter across all catalogues sources and the entire $V_{\rm circ}$ range was found to be 12.5\%, which is typical of our other uncertainties and in line with similar comparisons with other catalogues \citep{Courteau1997}.
Complete statistics for the RMSE of the scatter for each catalogue are shown in \Table{vcirc_stats}.
No catalogues were identified as significant outliers in the scatter analysis.
Consequently, no offsets or corrections to kinematics from different catalogues were required or applied, other than using a weighted average.
We note that the $V_{\rm circ}$ offsets in \Fig{vcirc_comp} for SHIVir and ACSVCS are typically negative for massive galaxies (where ETGs are more prevalent), while the offsets for \atlas galaxies are typically positive over the same mass range.
This is likely due to the fact that velocities for the SHIVir and ACSVCS galaxies were extracted from longslit spectra using \pPXF by us, whereas the \atlas values were extracted by others from IFU spectra.
The different fields-of-view of a slit relative to an IFU could impart an additional bias.

\begin{table}
\begin{center}
{\small
\begin{tabular}{c|c|c|c}
\hline\hline
Catalogue & Nb of meas. & Mean of $\frac{\Delta V_{\rm circ}}{V_{\rm circ, avg}}$ & RMSE of $\frac{\Delta V_{\rm circ}}{V_{\rm circ, avg}}$ \\
\hline									
ACSVCS & 72 & -0.7\% & 10.1\%\\
ALFALFA & 53 & -2.7\% & 13.2\%\\
\atlas & 42 & 6.0\% & 8.9\%\\
Chemin+03 & 5 & 4.7\% & 17.3\%\\
Fouqué+90 & 41 & -5.1\% & 20.0\%\\
Geha+03 & 14 & 3.8\% & 9.0\%\\
Rubin+97 & 13 & 5.2\% & 20.9\%\\
Rys+14 & 7 & -15.7\% & 20.6\%\\
SHIVir & 126 & 1.3\% & 11.0\%\\
SMACKED & 29 & 1.7\% & 9.9\%\\
van Zee+04 & 12 & -9.6\% & 19.6\%\\
\hline
All & 414 & -0.06\% & 12.5\%\\
\hline\hline
\end{tabular} }
\caption{Statistics of the relative scatter, $\Delta V_{\rm circ}$ / $V_{\rm circ, avg}$, for circular velocities from different catalogue sources, as a function of the weighted average $V_{\rm circ,avg}$ calculated for all available catalogues as plotted in \Fig{vcirc_comp}.
See \se{add_data} for references to the different catalogues.}
\label{tab:vcirc_stats}
\end{center}
\end{table}


\section{Spectroscopic Sample Parameter Distribution}
\label{sec:param_dist}

Photometric parameters are available for the entire SHIVir catalogue (742 galaxies, see \se{Photometry}), and can be matched with our ``SpecSHIVir'' sub-catalogue which consists of 190 SHIVir galaxies with spectroscopic data (O17).  
\Fig{Photo_Hist} shows the distributions of effective radius, $R_{\rm e}$; the 23.5 \magarc level isophotal radius, $R_{23.5}$; $i$-band absolute magnitude, $M_{i}$; $g-i$ colour; concentration, $C_{28}$; effective $i$-band surface brightness, $\mu_{\rm e}$; distance, $d$; and inclination, $i$; 
for both the full SHIVir (black) and SpecSHIVir (red) catalogues. 
Isophotal fitting failed for a few galaxies resulting in unphysical values for $R_{23.5}$, $g-i$, $C_{28}$, and $\mu_{\rm e}$. 
In these cases, the parameter in question was left out of the histogram. 
A distance to VCGs of $d = 16.5$ Mpc was assumed for galaxies without a measured distance (see \se{distances}); 
this explains the bump in the distance distribution at 16.5 Mpc.
The distance of 16.5 Mpc, taken as the core distance to the Virgo cluster \citep{Mei2007}, is indicated by a blue dashed line in \Fig{Photo_Hist}.

Radial distributions show mostly Gaussian functions with extended wings at the low-end due to the log-scale. 
In both the $R_{\rm e}$ and $R_{23.5}$ cases, the SpecSHIVir distributions are roughly Gaussian-like with centres at ${\sim}2$ kpc and ${\sim}3$ kpc respectively. 
While our spectroscopic sub-catalogue covers a fair range of galaxy sizes, there is a clear bias towards larger sizes. 
The Kormendy relation \citep{Kormendy1977,Hamabe1987}, or a linear trend between $\mu_{\rm e}$ and log $R_{\rm e}$, suggests a similar bias towards higher SB galaxies in our SpecSHIVir distribution of $\mu_{\rm e}$. 
Note that the distributions of \emph{all} galaxies, both LTGs and ETGs, are displayed. 
If an interesting distribution, such as the bimodality in SB found in \citet{McDonald2009b} and O17, were to be present for galaxy morphologies (or colours, concentrations, or other classifying criteria) shown separately, it would be hidden in \Fig{Photo_Hist}. 
The bias towards larger and brighter systems is again seen in the histogram of $M_{i}$. 
Bright galaxies not observed spectroscopically were omitted for practical reasons, including: LTGs with especially low ($< 20^{\circ}$) or high ($> 70^{\circ}$) inclinations, galaxies with major axes exceeding the slit length of our spectrographs, galaxies harboring active galactic nuclei (AGNs) or low-ionisation nuclear emission-line regions (LINERs), etc.

Both $g-i$ colour and $C_{28}$ concentration, defined in \Eq{c28}, can be used as a proxy for morphological classification. 
\Fig{SHIVir_Morphology} shows that the full SHIVir catalogue (or VCC/SDSS) is representative of the morphologies found in the VCC. 
Here again, \Fig{Photo_Hist} shows that the SpecSHIVir sample covers nearly the entire range of colour, with $g-i = 0-1.5$, and concentration with $C_{28} = 1.5-6$.
Similarly, the inclinations of our systems range from $0^{\circ}$ to almost $90^{\circ}$, the great majority lying between $20^{\circ}$ and $70^{\circ}$ as specified by our spectroscopic catalogue selection.
Systems with inclinations higher than $70^{\circ}$ or lower than $20^{\circ}$ are mostly ETGs, for which inclination plays little role and was not considered for their sample selection. 
Finally, the distribution of distances is quite wide, from 6 Mpc to 33.5 Mpc, with a significant peak around 16.5 Mpc.  
We can also appreciate the aforementioned issues with accurate distance assessments for VCGs: it is almost certain that part of the SHIVir sample is found in the W and W$^{\prime}$ groups, centred at ${\sim}35$ Mpc and ${\sim}30$ Mpc respectively \citep{Tully2016}. 
Distance uncertainties and group contamination as possible sources of scatter in the TFR were discussed O17.


\section{Kinematic Profiles}
\label{sec:kinprofiles}
\subsection{Rotation Curves}

\begin{figure}
\centering
\includegraphics[width=0.45\textwidth]{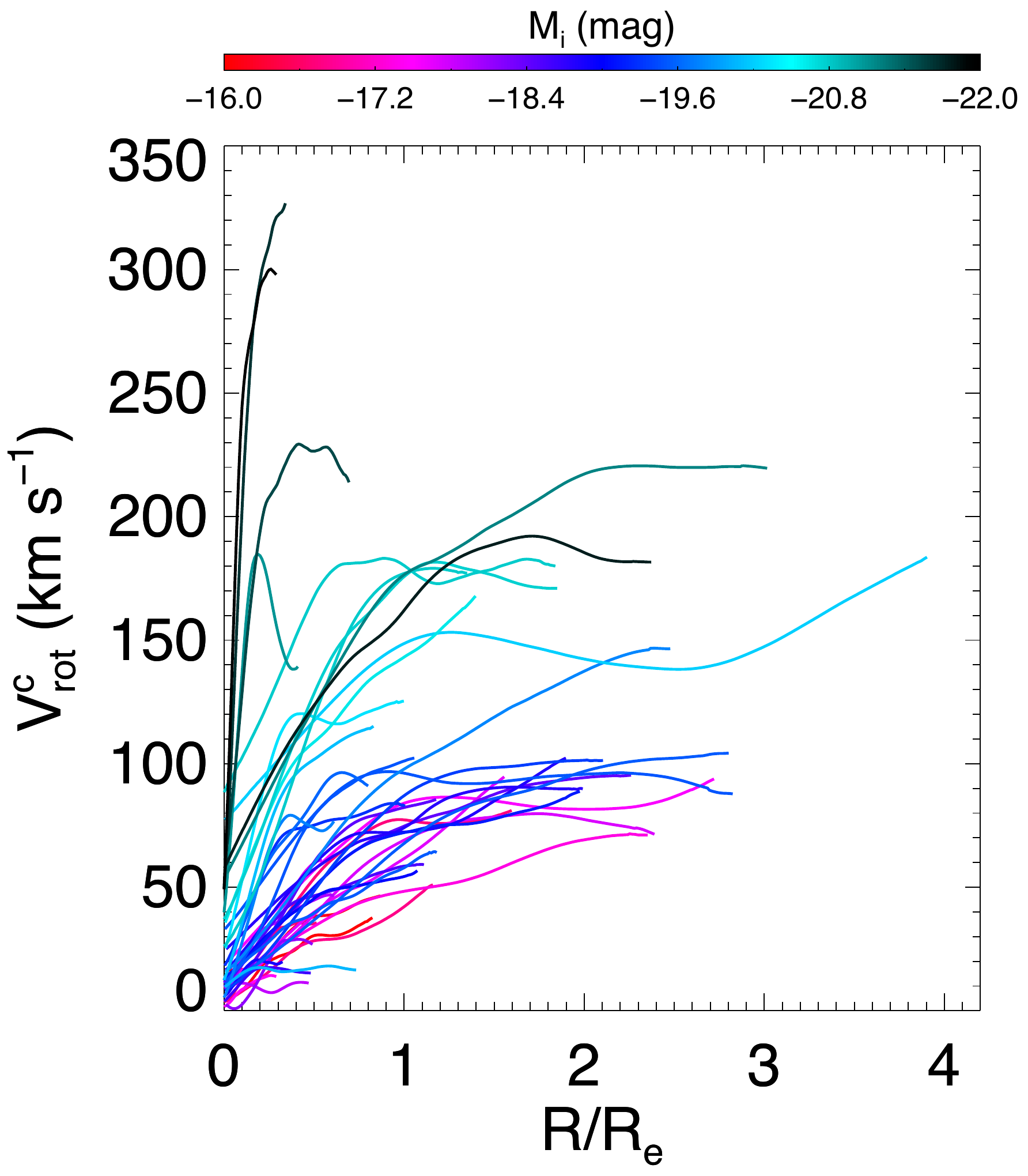}
\caption{Stack of 46 SHIVir inclination-corrected RCs showcasing profile shapes. 
Colour indicates the absolute magnitude $M_{i}$ of the galaxy.}
\label{fig:LineRC_Stack}
\end{figure}

We have presented in \se{rotvel} our method of extracting RCs from long-slit spectroscopy.  
RCs were modeled with a multi-parameter fitting function (see \Eq{mpf})
in order to map a rising slope, a turnover point (specified by a turnover radius and a sharpness parameter), and a mostly flat regime (specified by a maximum rotational velocity).
This fitting function also accounts for declining RCs, but this feature was never implemented since falloffs in profile's outskirts are rarely seen in cluster \ha\ RCs due to the limited extent of the emitting gas. 
Note also that SHIVir SB profiles extend out to $> 10 \ R_{\rm e}$, whereas SHIVir RCs typically reach $1-2\  R_{\rm e}$. 
A stack of our extracted inclination-corrected RCs is shown in \Fig{LineRC_Stack} (individual plots of all SHIVir RCs are found in the supplementary material). 
The RCs have been colour-coded by absolute magnitude. 
In accordance with our TFR analysis (see O17, \S~3.3), this figure verifies that brightest galaxies rotate fastest. 

\subsection{Velocity Dispersion Profiles}

\begin{figure*}
\centering
\includegraphics[width=0.99\textwidth]{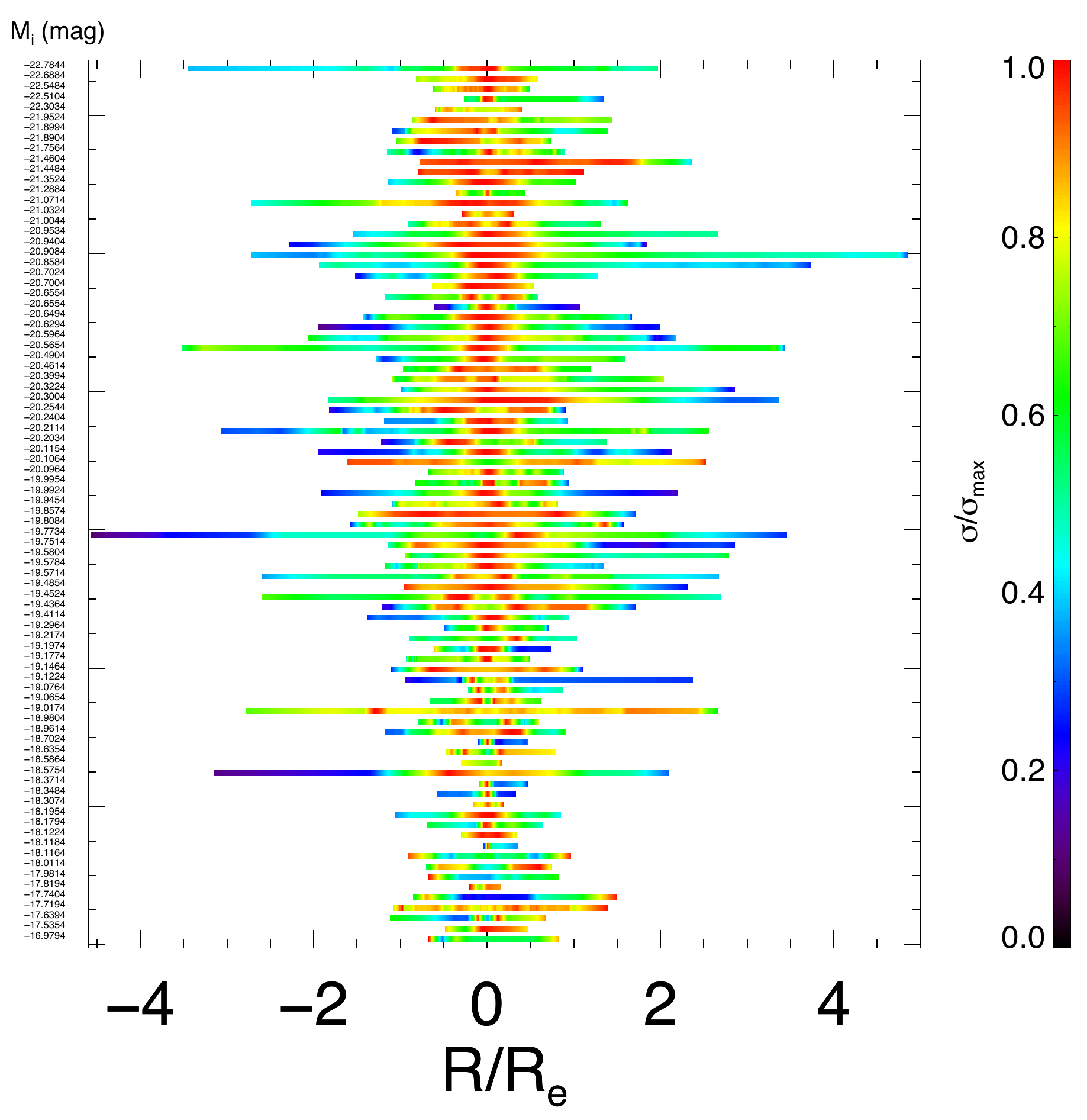}
\caption{Representation of 85 SHIVir VD profiles ordered by magnitude. 
VDs have been normalised by the maximum VD $\sigma_{\rm max}$ for each galaxy. 
The $i$-band absolute magnitude for each galaxy is shown on the left.}
\label{fig:ColourVD_Stack}
\end{figure*}

\begin{figure}
     \centering
     \subfigure[Example of a sigma-drop VD profile for VCC 759.]{
          \label{fig:sigmadrop_ex}
          \includegraphics[width=0.22\textwidth]{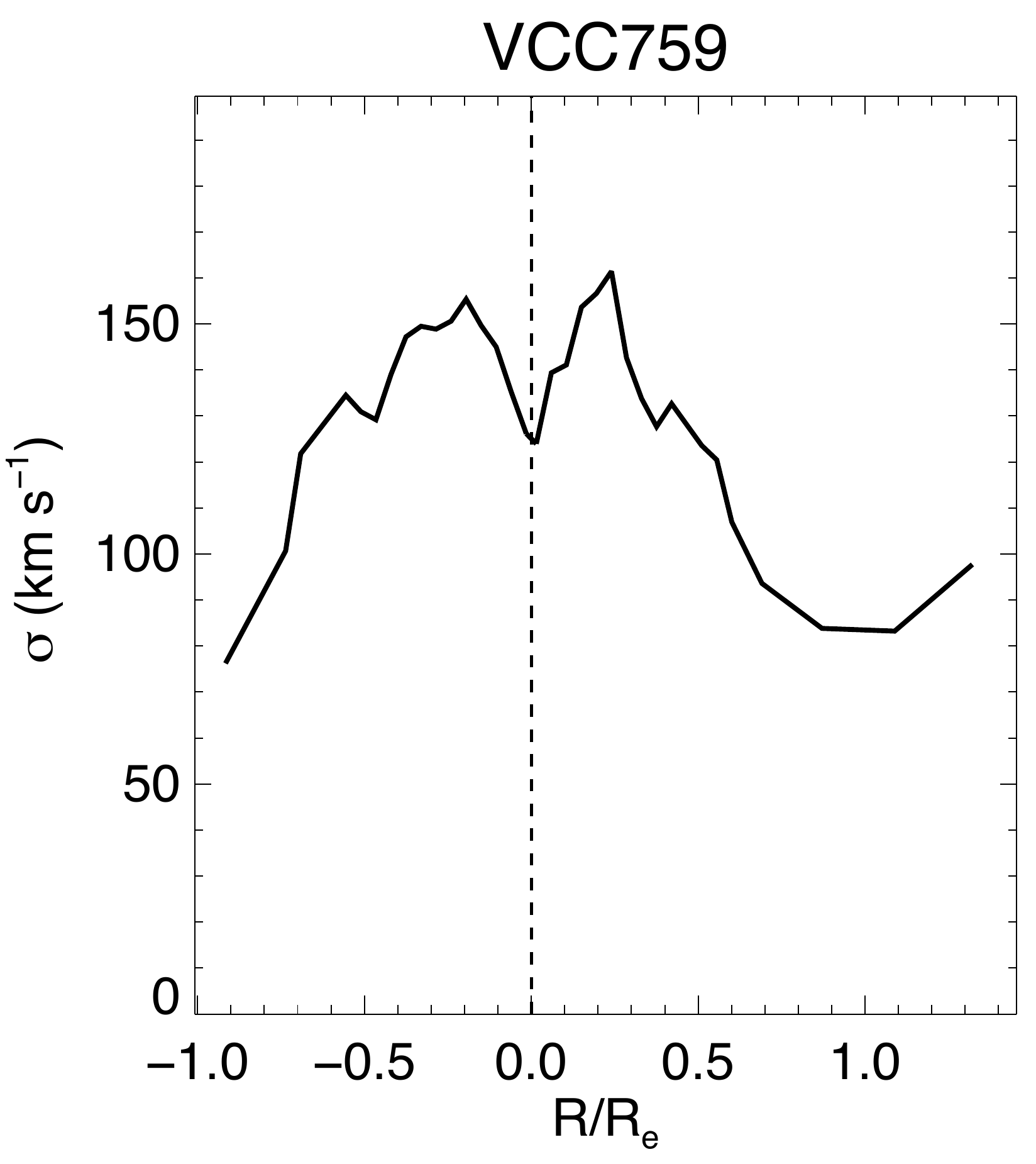}}
     \subfigure[Example of a rising VD profile for VCC 1528.]{
          \label{fig:rising_ex}
          \includegraphics[width=0.22\textwidth]{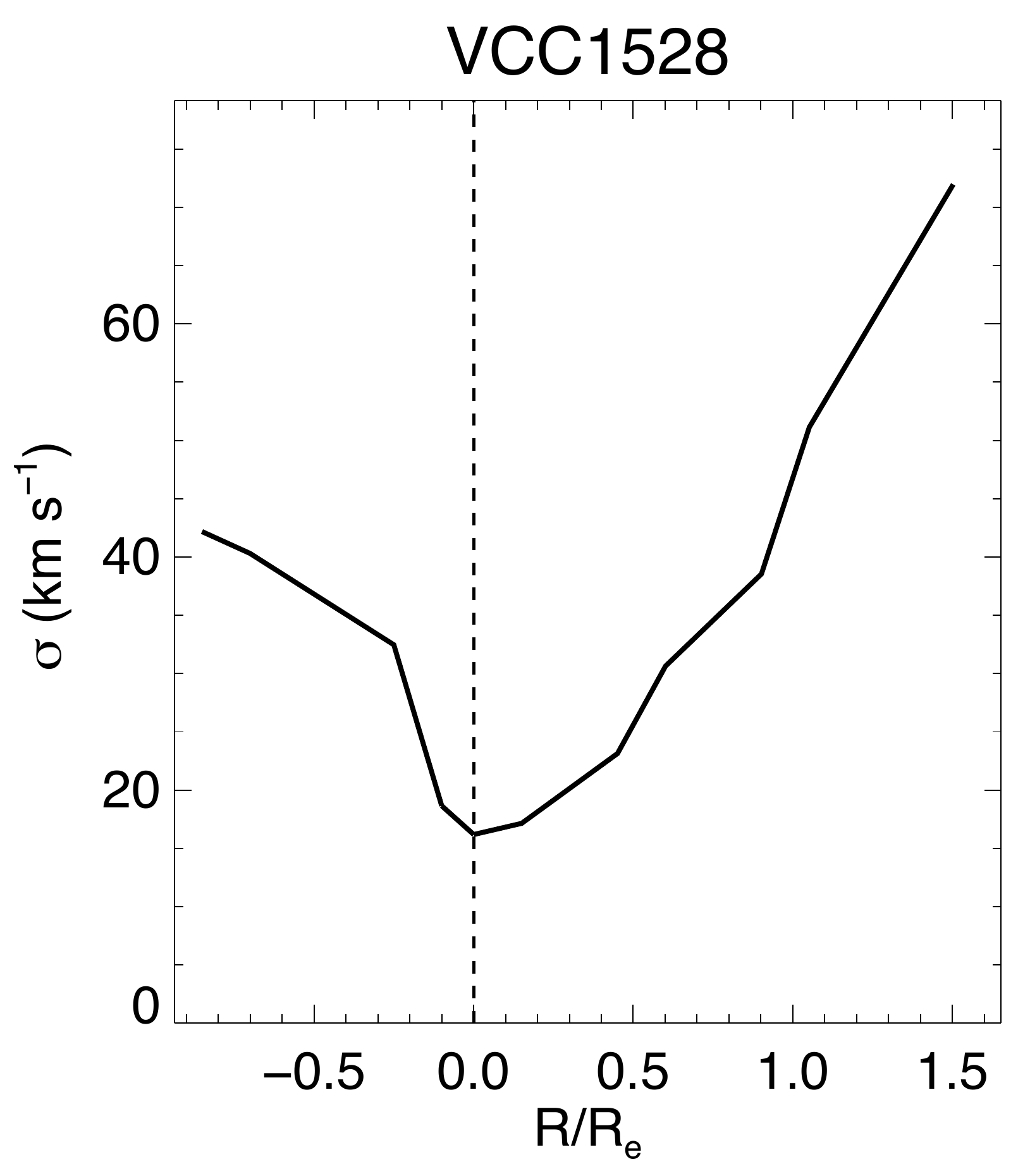}}\\
     \vspace{.2in}
     \subfigure[Example of a flat VD profile for VCC 1316.]{
           \label{fig:flat_ex}
           \includegraphics[width=0.22\textwidth]{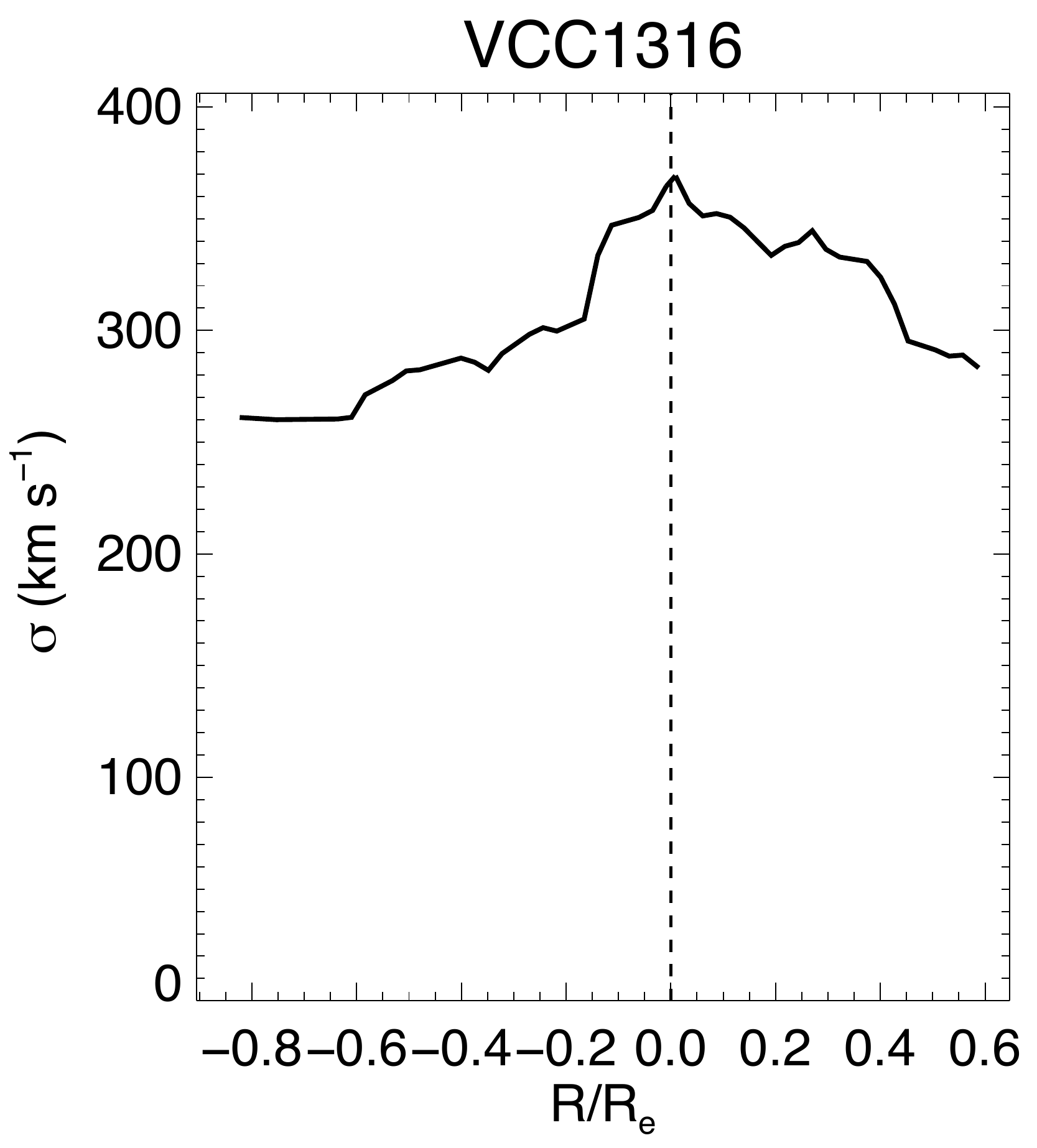}}
     \subfigure[Example of a peaked VD profile for VCC 958.]{
           \label{fig:peaked_ex}
          \includegraphics[width=0.22\textwidth]{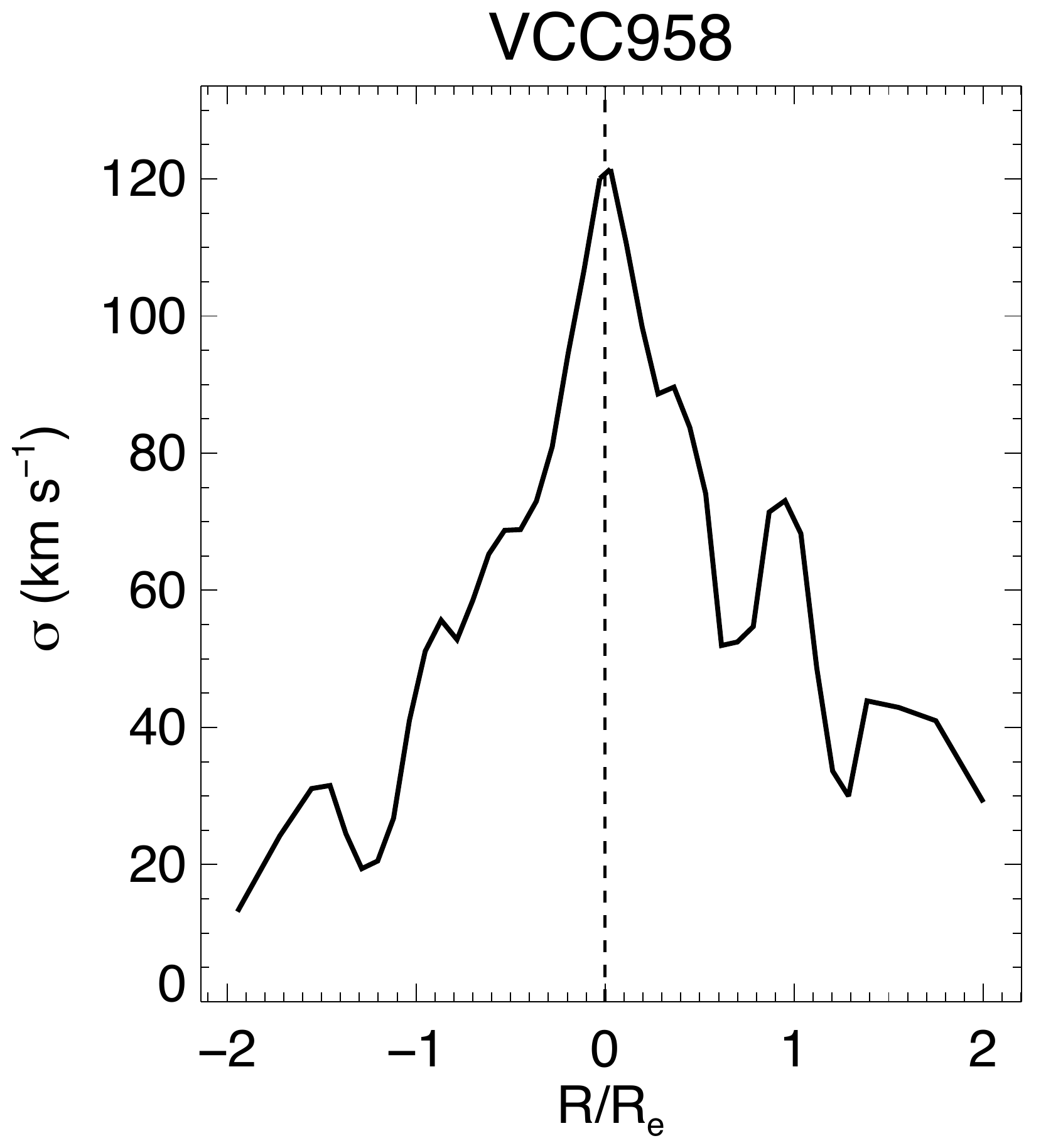}}
     \caption{Examples of VD profiles classified as: (a) sigma-drop, (b) rising, (c) flat, and (d) peaked.}
     \label{fig:VDprof_ex}
\end{figure}

\begin{table*}
\begin{center}
{\small
\begin{tabular}{c|cc|cc|ccc|cc|cccc}
\hline\hline
VD Profile Shape			& r/rs	& s		& B/AB	& A		& E		& S0		& Other	& Dwarf	& Giant	& $M_{*,{\rm bin} 1}$	& $M_{*,{\rm bin} 2}$ 	& $M_{*,{\rm bin} 3}$	& $M_{*,{\rm bin} 4}$ 	\\
\hline
Sigma-drop (24\%)		& 55\%	& 45\%  	& 50\%	& 50\%	& 20\%	& 55\%	& 25\%	& 5\%	& 95\%	& 10\%	& 40\%	& 25\%	& 25\%	\\
Rising (7\%)			& 0\%	& 100\%  	& 0\%	& 100\%	& 67\%	& 0\%	& 33\%   	& 67\%	& 33\%	& 83\%	& 0\%	& 0\%	& 17\%	\\
Flat (26\%)			& 17\%	& 83\%  	& 22\%	& 78\%	& 35\%	& 57\%	& 8\%   	& 26\%	& 74\%	& 41\%	& 27\%	& 5\%	& 27\%	\\
Peaked (44\%)			& 21\%	& 79\%  	& 37\%	& 63\%	& 29\%	& 55\%	& 16\%   	& 13\%	& 87\%	& 19\%	& 24\%	& 43\%	& 14\% 	\\
\hline
\hline
\end{tabular} }
\caption{Distribution of galaxies by profile shape classified as: (1) ringed (r/rs) or unringed (s), (2) barred (B/AB) or unbarred (A), (3) elliptical (E), lenticular (S0), or another morphology, (4) dwarf or giant, (5) in a certain stellar mass bin.
The stellar mass bins are: (1) ${\rm log}(M_{*}) < 9.5$, (2) $9.5 \leq {\rm log}(M_{*}) < 10.0$, (3) $10.0 \leq {\rm log}(M_{*}) < 10.5$, and (4) ${\rm log}(M_{*}) \geq 10.5$. 
The percentage of galaxies with a certain profile shape is indicated in parentheses next to the profile shape name.}
\label{tab:prof_class}
\end{center}
\end{table*}

\begin{table}
\begin{center}
{\small
\begin{tabular}{c|cccc}
\hline\hline
Galaxy Type/Feature		& Peaked	& Flat	& Rise	& Sigma-drop	\\
\hline
E (31\%)				& 42\%	& 31\%	& 15\%	& 12\%		\\
S0 (51\%)				& 48\%	& 30\%	& 0\%	& 22\%		\\
Other (18\%)			& 40\%	& 13\%	& 13\%	& 34\%		\\
\hline
r (16\%)				& 21\%	& 21\%	& 0\%	& 58\%		\\
rs (9\%)				& 50\%	& 12\%	& 0\%	& 38\%		\\
s (75\%)				& 48\% 	& 30\%	& 10\%	& 12\%		\\
\hline
B (26\%)				& 50\%	& 18\%	& 0\%	& 32\%		\\
AB (7\%)				& 50\%	& 17\%	& 0\%	& 33\%		\\
A (67\%)				& 42\%	& 32\%	& 11\%	& 16\%		\\
\hline
Dwarf (20\%)			& 31\%	& 38\%	& 25\%	& 6\%		\\
Giant (80\%)			& 48\%	& 25\%	& 3\%	& 24\%		\\
\hline
$M_{*,{\rm bin} 1}$ (27\%)		& 30\%	& 39\%	& 22\%	& 9\%		\\
$M_{*,{\rm bin} 2}$ (27\%)		& 39\%	& 26\%	& 0\%	& 35\%		\\
$M_{*,{\rm bin} 3}$ (26\%)		& 73\%	& 5\%	& 0\%	& 23\%		\\
$M_{*,{\rm bin} 4}$ (20\%)		& 29\%	& 35\%	& 6\%	& 29\%		\\
\hline
\hline
\end{tabular} }
\caption{As in \Table{prof_class}, but with columns and rows inverted.}
\label{tab:morph_class}
\end{center}
\end{table}

The extraction of integrated VDs and VD profiles was presented in \se{vd_extr}. 
Attempts to study and model VD profiles \citep{Gerhard2001,Battaglia2013} and maps \citep{Krajnovic2011,Forbes2016} as a function of global galaxy parameters have been made (see Fig.22 of \citealp{Cappellari2016}), often in conjunction with the study of mass distribution and assembly history \citep{Chae2014,Forbes2016}. 
Unlike LTGs for which RC amplitude and shape are mostly correlated with the absolute luminosity of the galaxy through the TFR, ETGs are more complex likely as a result of their comparatively more evolved formation histories. 
There is for instance greater diversity in the shape of ETG kinematic profiles, relative to LTGs. 
In an attempt to establish links between profile shapes and structural parameters for our SHIVir ETGs, we have displayed all 85 resolved VD profiles in \Fig{ColourVD_Stack}. 
Note that a construction similar to \Fig{LineRC_Stack} for VDs is not possible due to the varied nature of the ETG VD profile shapes. 
For the SHIVir VD profiles, colour indicates VD normalised by the maximum VD of a profile $\sigma_{\rm max}$ (individual plots of all SHIVir VD profiles are found in the supplementary material). 
The profiles have been ranked by absolute magnitude, with the brightest galaxies at the top. 
A bright red region at the centre 
of the profile signifies a strong peak, while a green or blue region at the centre indicates a trough. 
The percentage of profiles with central troughs is much higher for fainter galaxies. 
This implies that strongly peaked and radially decreasing profiles are characteristic of more massive ETGs, whereas dwarf galaxies have more complex profiles with frequent central depressions.

Our ETG profiles are classified into four categories: sigma-drop, rising, flat, and peaked. 
The first is a peaked profile with a small trough at the centre, while the second has its lowest $\sigma$ value at the centre and is mostly rising with increasing radii. 
The third and fourth are self-explanatory. 
Examples of all four profile shapes are shown in \Fig{VDprof_ex}. 
The sigma-drop signature can be mistaken for the so-called ``2$\sigma$'' feature \citep{Krajnovic2011} which is commonly thought to be the signature of a kinematically-decoupled core. 
The peaks in a 2$\sigma$ feature are usually separated by a larger distance than in a sigma-drop profile; \citet{Krajnovic2011} set this distance to be at least $0.5 R_{\rm e}$. 
Here, we do not differentiate between a sigma-drop and a 2$\sigma$ feature, but identify any known kinematically-decoupled cores in the notes accompanying all resolved VD profiles provided in the supplementary material.

The profile shape classification was performed visually. 
Certain galaxies whose profile shapes are a combination of multiple categories could be found; an admittedly subjective choice (not unlike those applied for morphological typing of galaxies) was made regarding the more dominant shape in these cases. 
Fortunately, our ETG sample of 85 VD profiles should mitigate the impact of borderline cases.  
With those caveats in mind, we find that 24\% of our galaxies have sigma-drop profiles, 7\% have rising profiles, 26\% have flat profiles, and 44\% have centrally peaked profiles.

The link between VD profile shapes and structural/morphological components of galaxies has been previously explored for bulge signatures \citep{Fabricius2012} and nuclear rings and discs in sigma-drop profiles \citep{Comeron2010}. 
With our SHIVir kinematic sample, we can determine the occurrence of certain profile shapes in association with a number of morphological characteristics: rings, bars, morphological type, and size. 
These characteristics were taken from the NASA Extragalactic Database\footnote{Operated by the Jet Propulsion Laboratory, California Institute of Technology, under contract with the National Aeronautics and Space Administration.}. 
The breakdown of profile shapes was also determined for four different stellar mass bins: (1) ${\rm log}(M_{*}) < 9.5$, (2) $9.5 \leq {\rm log}(M_{*}) < 10.0$, (3) $10.0 \leq {\rm log}(M_{*}) < 10.5$, and (4) ${\rm log}(M_{*}) \geq 10.5$. 
\Table{prof_class} and \Table{morph_class} present our sample according to these characteristics and profile shapes. 
We find the following:

1) Most galaxies containing inner rings produce sigma-drop profiles, though many sigma-drop profiles are also found in unringed galaxies. 
\citet{Comeron2010} theorised that the inflow-induced star formation likely feeding the ring may also cause the sigma-drop kinematic signature. 
However, we find that rings are only one of several mechanisms that are associated with sigma-drop profiles, all of which are expected to be located in the innermost region of the galaxy. 
Other possible progenitors include nuclear dust spirals and Seyfert activity \citep{Comeron2010}.
\citet{Koleva2008} cautioned that the appearance of the sigma-drop signature may be induced by a mismatch in the stellar template's metallicity during the VD extraction. 
Some sigma-drop profiles may also be due to kinematically-decoupled cores. 
A 2D kinematic map would enable a more definitive classification.

2) Dwarf (or low mass) galaxies do not typically produce sigma-drop profiles.

3) The majority of peaked profiles are found in giant (i.e. not classified as dwarfs), mostly unringed, unbarred galaxies with $M_{*} > 10^{10} \solarm$.
This result is mirrored in \Fig{ColourVD_Stack}. 
These systems are the most classical, featureless ETGs and are almost purely pressure-supported.

4) Rising dispersion profiles are found in four dwarf Es and two early-type spirals; the spirals also show detectable RCs. 
Five out of these six galaxies with rising dispersion profiles were also found to have $M_{*} \lt 9.5 \solarm$. 
A substantial fraction of the kinetic energy for these galaxies  is also likely found in their rotational, rather than dispersion, component. 

No direct correlation between profile slope and morphology or concentration was found, but this may be due in part to the complex profile shapes for which a single slope is not easily identified.  
Moreover, except for the rising profile type, we do not find a strong correlation between profile shapes and our mass bins. 
This is likely due to the small number of low-mass galaxies in our sample, and that above a stellar mass threshold of approximately $10^{10} \solarm$, stellar mass becomes secondary to other galaxy parameters for predicting dispersion profile shape.

\begin{figure}
\centering
\includegraphics[width=0.45\textwidth]{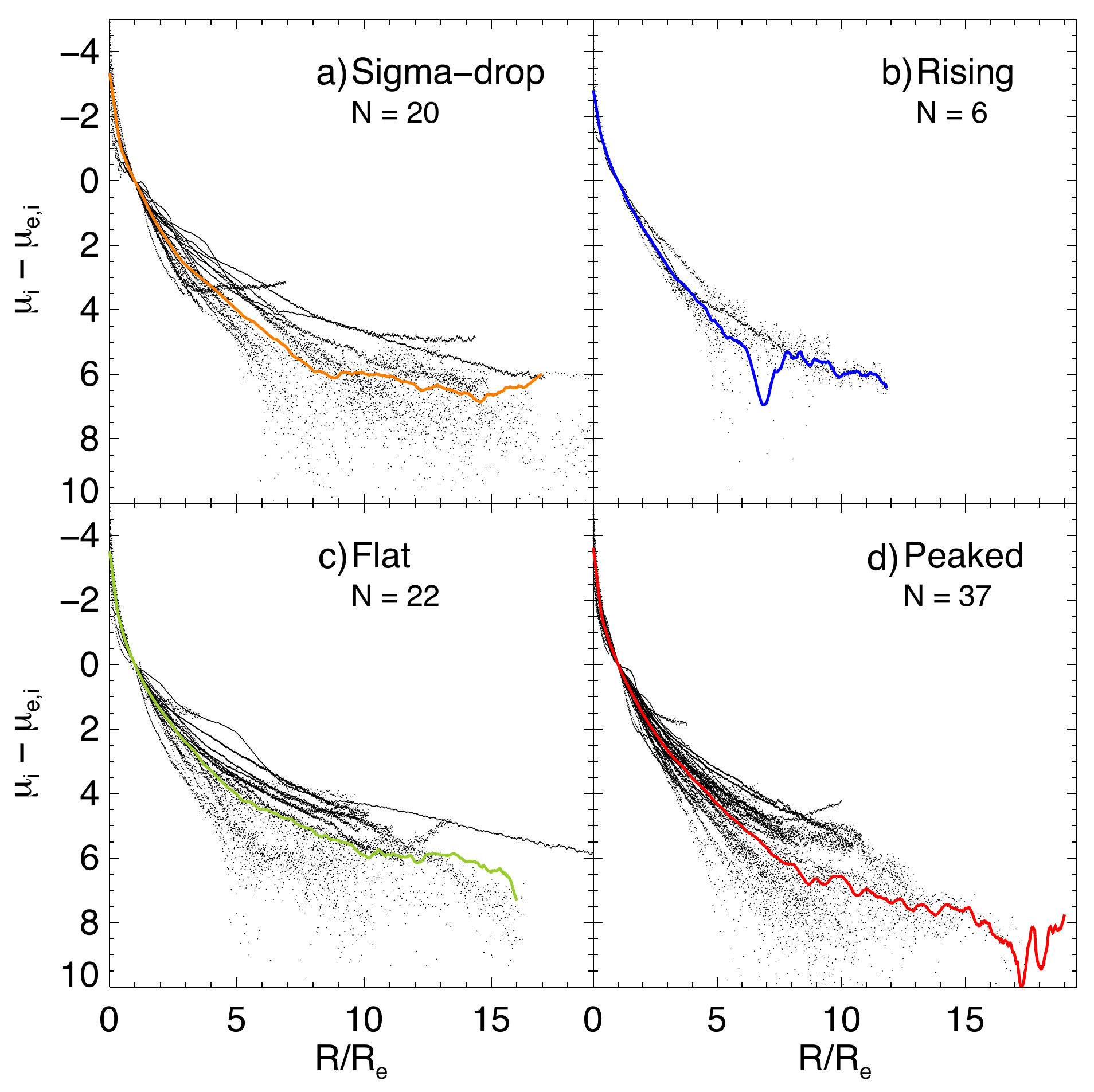}
\caption{SB profiles binned by kinematic profile shape: a) sigma-drop, b) rising, c) flat, and d) peaked}.  
The average SB profiles for each kinematic shape bin are shown in colour. 
The number of profiles in each shape bin $N$ is indicated.
\label{fig:SBProf_Class}
\end{figure}

\begin{figure}
\centering
\vspace{.15in}
\includegraphics[width=0.45\textwidth]{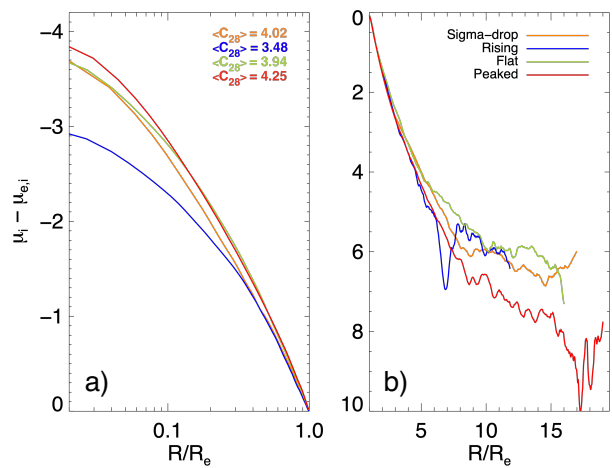}
\caption{Comparison of average SB profiles by kinematic shape bin, for a) $R < R_{\rm e}$, and b) $R > R_{\rm e}$. 
The average $i$-band concentration $<C_{28}>$ for each kinematic shape bin is also shown.}
\label{fig:AvSB_Prof}
\end{figure}

\Fig{SBProf_Class} displays our sample's normalised SB profiles by kinematic profile shape in order to highlight any commonalities due to galaxy kinematics. 
The average SB profiles for each shape bin are also plotted in \Fig{AvSB_Prof}, for both $R < R_{\rm e}$ and $R > R_{\rm e}$. 
A similar study of SHIVir SB profiles, but for concentration and morphology, was provided by \citet{McDonald2011}. 
At small radii, galaxies with rising VD profiles appear more cored, usually associated with lenticulars containing pseudobulges \citep{MacArthur2003,Kormendy2004}. 
These also have low concentrations, with an average $C_{28} = 3.5$; for comparison, structures with $C_{28} = 3.8$ or less are typically associated with pseudobulges (M09). 
Galaxies with rising VD profile shapes also appear to have shallower SB profiles that terminate at smaller radii than for any other kinematic shape. 
Conversely, galaxies with peaked profiles have more extended SB profiles, in addition to high concentrations (average of $C_{28} = 4.5$) and the cuspiest inner SB profiles. 
This confirms our previous assessment that peaked VD profiles occur predominantly in massive ETGs.  
The comparison of all four averaged SB profiles in \Fig{AvSB_Prof} shows great commonality, yet significant departure from those of late-type spirals presented in Fig.~11 of \citet{McDonald2011}.


\section{Summary \& Conclusion}
\label{sec:Conclusion}

Understanding the local and external drivers of galaxy evolution is a thriving subfield of astrophysics. 
Along those lines, our SHIVir sample offers a deep look at the Virgo cluster of galaxies by maximising the power of long-slit spectroscopy to assemble a dynamical catalogue of Virgo galaxies. 
We have presented a comprehensive overview of the spectroscopic component of the SHIVir survey, including an extensive dynamical database and set of kinematic profiles. 
Our current SHIVir spectroscopic sample contains 190 Virgo cluster galaxies; novel spectroscopic observations were acquired for 138 of these, and kinematic information for the remaining 52 galaxies was taken from select literature sources.
Dynamical masses were derived for the entire sample and coupled with photometrically determined stellar mass measurements.
From our novel spectroscopic observations, \ha\ rotation curves and resolved velocity dispersion profiles were produced for 46 and 85 SHIVir galaxies, respectively. 
A total of 16 galaxies have both RC and VD profiles available. 
An analysis of the scaling relations and stellar-to-halo-mass relations based on this catalogue was presented in O17.

We found the overall RC shapes of Virgo galaxies to match those of galaxies residing in other environments, with the notable exception that they are typically truncated in both \ha\ and \ion{H}{I}. 
This is likely due to environmentally-dependent processes such as ram-pressure stripping, tidal stripping and interactions. 
In spite of this difference, the velocity-stellar mass relation of VCGs is largely unaffected by environment (i.e., same TFR slopes) as shown in O17.  
All agents responsible for the truncated kinematic tracers apparently do not influence the dark matter halo which strongly regulates a spiral galaxy's maximal rotational velocity, but may simply increase the scatter in cluster TFRs.

We paid close attention to VD extraction and offer various recommendations regarding the use of \texttt{pPXF} for the acquisition of stellar VDs: 
1) spectra with SNR greater than 50/\AA\ are required to keep the relative error below 5\%; 
2) a large stellar template library is needed for optimal spectral fitting, or else using a custom library for each galaxy; and,
3) carefully-adjusted emission line mask widths are crucial, especially for galaxies with $\sigma < 100 \kms$ and galaxies with known ionised gas. 

Resolved VD profiles were also studied and classified to determine potential links between kinematic signatures and morphological/structural parameters. 
Rings were identified as one of many potential drivers for sigma-drop profiles. 
Rising VD profiles were detected in mostly dwarf (low-mass) elliptical and lenticular galaxies, whereas flat, peaked, and sigma-drop profiles all appeared in giant (high-mass) galaxies.

The distribution of $i$-band magnitude $M_{i}$ and effective radius $R_{\rm e}$ shows that our SHIVir sample is, unsurprisingly, biased towards larger and brighter targets as high SNR data are more easily obtained for them. 
Nonetheless, our sample covers a representative range of other parameters such as inclination $i$, $g-i$ colour, and concentration $C_{28}$ over the SDSS footprint of the Virgo cluster. Augmenting the lower-mass range of SHIVir would more accurately represent the galaxy population of the Virgo cluster. 

\subsection{Future Work}

The SHIVir photometric and spectroscopic catalogue of Virgo Cluster galaxies is meant to support studies of the Virgo cluster and galaxy structure in general (e.g., O17).  
Various steps may still be taken to augment its scope and reduce interpretation biases towards studying galaxy formation and evolution based on kinematic profiles and scaling relations.  
Such steps include: 

1) a systematic study of differences between kinematic tracers extracted from long-slit and IFU spectra \citep{Cappellari2011a,Arnold2014,Foster2016}, and refinement of aperture corrections for integrated VD measurements. 
Improved and more extended kinematics may further decrease the scatter of scaling relations, such as those presented in O17;

2) a more detailed VD profile classification to bolster the connections between kinematic signatures and structural/photometric parameters in ETGs. 
Our initial assessment of such links has relied on visual classification into four profile types. 
The shape, slope or variability of these profiles ought to be quantified in a more quantitative manner despite the complex profile variations; 

3) an extended dynamical database for low-mass galaxies in order to anchor O17's scaling relations in the low-mass regime where larger scatter prevails given a noticeable dearth of data, particularly for LTGs;

4) stellar population parameters such as age and metallicity to further isolate population trends, reduce the scatter of the Virgo cluster scaling relations, and better understand how star formation and feedback affect mass aggregation. 
Age, metallicity, and kinematics can all be determined from optical and near-infrared spectroscopy. 
Moreover, expected trends between profile kinematic signatures and stellar population parameters may be identified; and, 

5) full integration of the NGVS \citep{Ferrarese2020}, DESI \citep{Dey2019}, and SHIVir survey data. 

Other long-term endeavours should include mapping VCGs with efficient wide FOV IFUs in order to collect 3D spectrophotometric data in single observations.  
Such FOVs must be considerably larger than those of currently available IFUs and would enable mapping large VCGs beyond $R_{\rm e}$ (smaller systems would be fully sampled). 
The choice of a kinematic and/or photometric position angle would no longer be a limitation.

An expanded velocity catalogue would also enable the construction of the Virgo Velocity Function.
The number of galaxies per unit volume as a function of halo circular velocity, $N(V)$, is a major prediction of cosmological models, particularly at low masses where $N(V)$ sets stringent constraints on $\Lambda$CDM models \citep{Klypin1999,Papastergis2011,TrujilloGomez2011}. 
The SHIVir catalogue can be used to derive a direct, homogeneous galaxy circular velocity function for both gas-rich and gas-poor galaxies following a careful completeness correction, in addition to determining the variations of galaxy dynamical properties over the multivariate parameter space of velocity, luminosity, radius, and colour.


\section*{Acknowledgments}

NNQO and SC are grateful to the Natural Sciences and Engineering Research Council of Canada, the Ontario Government, and Queen's University for support through various generous scholarships and grants. 
NNQO also acknowledges support from the Trottier Family Foundation.
We wish to thank Nikhil Arora for enlightening discussions, as well as Enrico Maria Corsini and Lorenzo Morelli for the Padova templates used in our spectral fitting. 
Our referee also provided constructive comments and suggestions that improved the presentation of this paper for which we are thankful.

Part of our data are based on observations at Kitt Peak National Observatory at NSF’s NOIRLab (NOIRLab Prop. IDs 06A-0538, 07A-0477; PI: S. Courteau), which is managed by the Association of Universities for Research in Astronomy (AURA) under a cooperative agreement with the National Science Foundation. The authors are honored to be permitted to conduct astronomical research on Iolkam Du’ag (Kitt Peak), a mountain with particular significance to the Tohono O’odham.

Part of our data are based on observations obtained with the Apache Point Observatory 3.5-meter telescope, which is owned and operated by the Astrophysical Research Consortium.

Part of our data are based on observations obtained at the international Gemini Observatory, a program of NSF’s NOIRLab [GS-2014A-Q-46; GS-2015A-Q-79], which is managed by the Association of Universities for Research in Astronomy (AURA) under a cooperative agreement with the National Science Foundation on behalf of the Gemini Observatory partnership: the National Science Foundation (United States), National Research Council (Canada), Agencia Nacional de Investigación y Desarrollo (Chile), Ministerio de Ciencia, Tecnología e Innovación (Argentina), Ministério da Ciência, Tecnologia, Inovações e Comunicações (Brazil), and Korea Astronomy and Space Science Institute (Republic of Korea).

\section*{Data Availability}

The data underlying this article are available in the online supplementary material, as well as 
online at \texttt{https://www.queensu.ca/academia/courteau/links}.

The ``Target List'', ``Kinematics'' and ``Masses'' tables are provided in CSV format along with a README file. 
The first 9 and last rows of each table along with the table header and caption are shown below for \Table{target_list}, \Table{kinematics} and \Table{masses}.
Additional plots (rotation curves and velocity dispersion profiles) along with detailed descriptions of individual VCGs' kinematics can also be found in the supplementary material in PDF format. 
Any use of these data should properly credit the authors of this paper.


\begin{table*}
\begin{center}
\begin{tabular}{cc|c@{$\ \pm$}ccc@{$\ \pm$}cc@{$\ \pm$}ccc@{$\ \pm$}cccc}
\hline\hline
VCC	& Other name	& \multicolumn2c{$m^{\rm c}_{i}$} 	& $\mu^{\rm i}_{{\rm e},i}$  & \multicolumn2c{$R^{\rm c}_{{\rm e},i}$}	& \multicolumn2c{$R^{\rm c}_{23.5,i}$}	& HT		& \multicolumn2c{$d$}		& $i$ 		& PA			& Obs. \\
--	& --	& \multicolumn2c{(mag)}	& (\magarc)	& \multicolumn2c{(kpc)}	& \multicolumn2c{(kpc)}	& --	& \multicolumn2c{(Mpc)}	& ($^\circ$)	& ($^\circ$)	& --\\
\hline
355	& NGC4262	& 10.7 	& 0.1		& 19.10	& 0.8 & 0.1		& 4.1 & 0.1	& 1			& 15.5 	& 0.35	& 28.9	& 115	& KPNO \\
389	& IC0781		& 12.7 	& 0.4 	& 22.53	& 1.5 & 0.3		& 2.9 & 0.4	& -3			& \multicolumn2c{--}	& 37.2	& 50		& APO \\
437 	& UGC07399A	& 13.1 	& 0.3 	& 22.78	& 1.5 & 0.1		& 3.7 & 0.1	& -1			& 17.2 	& 0.6		& 58.1	& 14		& KPNO \\
460 	& NGC4293	& 9.8 	& 0.2 	& 21.14	& 4.9 & 1.1		& 12.1 & 2.7 	& 3			& 14.1 	& 3.2		& 70.1	& --		& -- \\
490 	& IC0783		& 12.9 	& 0.5 	& 22.64	& 1.9 & 0.2		& 3.4 & 0.3	& -3			& 18.3 	& 1.5		& 35.6	& 146	& APO \\
510	& UGC07425	& 13.9 	& 0.9 	& 23.41	& 1.5 & 0.2		& 1.8 & 0.3	& -1			& \multicolumn2c{--}	& 37.7	& 127	& APO \\
522	& NGC4305	& 12.0 	& 0.2 	& 22.11	& 2.2 & 0.3		& 4.8 & 0.7	& 3			& \multicolumn2c{--}	& 60.5	& 61		& APO \\
523	& NGC4306	& 12.5 	& 0.3 	& 21.97	& 1.4 & 0.2		& 3.3 & 0.5	& -3			& \multicolumn2c{--}	& 44.6	& 129	& APO \\
543	& UGC07436	& 13.0 	& 0.5 	& 23.30	& 1.6 & 0.1		& 2.8 & 0.1	& -1			& 15.8 	& 0.6		& 58.2	& 153	& KPNO \\
... & ... & \multicolumn2c{...} & ... & \multicolumn2c{...} & \multicolumn2c{...} & ... & \multicolumn2c{...} & ... & ... & ...\\
2095	& NGC4762	& 9.8 	& 0.1 	& 20.75	& 3.7 & 1.4		& 14.8 & 5.5	& 1			& 16.3 & 6.1		& 82.1	& 62	& KPNO \\
\hline\hline
\end{tabular}
\caption{Target List. This target list of the 190 Virgo Cluster galaxies included in the “SpecSHIVir” subcatalogue tabulates the following parameters: (1) VCC designation, (2) Alternate designation, (3) Extinction-corrected apparent magnitude, (4) Effective $i$-band SB, extinction- and inclination-corrected, (5) Effective $i$-band radius, inclination-corrected, (6) Isophotal radius at the 23.5 mag arcsec$^{-2}$ level, inclination-corrected, (7) Hubble type, (8) Distance, (9) Inclination, (10) Position angle of the long-slit, if observed as part of novel SHIVir spectroscopic observations, (11) Site of observation, if observed as part of novel SHIVir spectroscopic observations.}
\label{tab:target_list}
\end{center}
\end{table*}


\begin{table*}
\begin{center}
\begin{tabular}{c|c@{$\ \pm$}cc@{$\ \pm$}cc@{$\ \pm$}cc@{$\ \pm$}cc@{$\ \pm$}cc@{$\ \pm$}cc}
\hline\hline
VCC	& \multicolumn2c{$V^{\rm c}_{2.2}$} 	& \multicolumn2c{$V^{\rm c}_{23.5}$}  	& \multicolumn2c{$\sigma_{0}$}	& \multicolumn2c{$\sigma_{R_{\rm e}/4}$}	& \multicolumn2c{$\sigma_{R_{\rm e}}$}	& \multicolumn2c{$\sigma_{2R_{\rm e}}$}	& Profile \\
 & \multicolumn2c{(km s$^{-1}$)}	& \multicolumn2c{(km s$^{-1}$)}		& \multicolumn2c{(km s$^{-1}$)}	& \multicolumn2c{(km s$^{-1}$)}		& \multicolumn2c{(km s$^{-1}$)}		& \multicolumn2c{(km s$^{-1}$)}		&  \\
\hline
355	& \multicolumn2c{--} 			& \multicolumn2c{--}			& 198 & 4	& 195 & 4	& 189 & 3	& 189 & 4	& VD \\
389	& \multicolumn2c{--}			& \multicolumn2c{--}			& 37 & 5	& 35 & 5	& 31 & 3	& 28 & 8	& VD \\
437 	& \multicolumn2c{--}			& \multicolumn2c{--}			& 21 & 8	& 22 & 6	& 34 & 13	& 33 & 11	& VD \\
490 	& \multicolumn2c{--}			& \multicolumn2c{--}			& 34 & 3	& 34 & 3	& 44 & 3	& 36 & 6	& VD \\
510	& \multicolumn2c{--}			& \multicolumn2c{--}			& 43 & 7	& 39 & 6	& 48 & 7	& 40 & 12	& -- \\
522	& \multicolumn2c{--}			& \multicolumn2c{--}			& 44 & 3	& 44 & 3	& 55 & 3	& 62 & 3	& VD \\
523	& \multicolumn2c{--}			& \multicolumn2c{--}			& 41 & 8	& 31 & 9	& 40 & 5	& 40 & 5	& -- \\
543	& \multicolumn2c{--}			& \multicolumn2c{--}			& 27 & 8	& 27 & 8	& 30 & 11	& 32 & 11	& -- \\
559	& 119 & 4					& 122 & 4					& 63 & 4	& 68 & 3	& 100 & 3	& \multicolumn2c{--}	& RC,VD \\
... & \multicolumn2c{...} & \multicolumn2c{...} & \multicolumn2c{...} & \multicolumn2c{...} & \multicolumn2c{...} & \multicolumn2c{...} & ... \\
2095	& \multicolumn2c{--}			& \multicolumn2c{--}			& 135 & 2	& 128 & 2	& 128 & 3	& \multicolumn2c{--}	& VD \\
\hline\hline
\end{tabular}
\caption{SHIVir Sample Kinematics. This table lists kinematics for the 138 Virgo Cluster galaxies with raw spectra: (1) VCC designation, (2) Rotational velocity at 2.15 disc scale lengths, (3) Rotational velocity at $R_{23.5,i}$, (4) Central velocity dispersion, (5) Velocity dispersion inside $R_{\rm e}/4$, (6) Velocity dispersion inside $R_{\rm e}$, (7) Velocity dispersion inside $2R_{\rm e}$, (8) Whether a rotation curve (RC) or velocity dispersion (VD) was extracted, or both. All rotational velocities are inclination-corrected.}
\label{tab:kinematics}
\end{center}
\end{table*}


\begin{table*}
\begin{center}
\def\arraystretch{1.4}
\begin{tabular}{c|c@{$\ \pm\ $}ccc@{$\ \pm\ $}ccc@{$\ \pm\ $}cc}
\hline\hline
VCC	& \multicolumn2c{$V_{\rm circ}$}	& $\log M_{*}$		& \multicolumn2c{$\log M_{\hi}$}	& $\log M_{\rm DM}$		& \multicolumn2c{$\log M_{\rm dyn}$}	& Lit. Sources \\
 	& \multicolumn2c{(\kms)}			& ($\log \solarm$) 	& \multicolumn2c{($\log \solarm$)}	& ($\log \solarm$)		& \multicolumn2c{($\log \solarm$)} & \\
\hline
355 			& 316 & 9     					& $10.17^{+0.09}_{-0.10}$      	& 8.69 & 0.01     				& $11.05^{+0.04}_{-0.05}$		& 11.10 & 0.02						& 1,2,3,4 \\
389 			& 57 & 5        					& $9.20^{+0.11}_{-0.11}$     		& \multicolumn2c{--}         			& $9.14^{+0.26}_{-0.26}$		       		& 9.47 & 0.04						& 1,10 \\
437 			& 75 & 14       					& $8.91^{+0.15}_{-0.10}$      	& \multicolumn2c{--}         			& $9.75^{+0.13}_{-0.12}$			& 9.81 & 0.08						& 1,2,10,11\\
460 			& 118 & 23        				& $10.23^{+0.18}_{-0.18}$		      	& 7.74 & 0.05      				& $10.34^{+0.43}_{-0.43}$		      		& 10.59 & 0.16						& 3,8 \\
490 			& 84 & 6        					& $9.21^{+0.13}_{-0.10}$      	& \multicolumn2c{--}         			& $9.77^{+0.10}_{-0.09}$       		& 9.88 & 0.03						& 1 \\
510 			& 92 & 13        					& $8.64^{+0.11}_{-0.10}$      	& \multicolumn2c{--}         			& $9.64^{+0.16}_{-0.15}$       		& 9.68 & 0.07						& 1 \\
522  			& 105 & 4        					& $9.58^{+0.13}_{-0.10}$      	& \multicolumn2c{--}         			& $9.92^{+0.08}_{-0.07}$      		& 10.09 & 0.02						& 1 \\
523 			& 80 & 8        					& $9.34^{+0.11}_{-0.12}$     	& \multicolumn2c{--}         			& $9.64^{+0.16}_{-0.16}$	 	     		& 9.81 & 0.05						& 1,9,10 \\
543  			& 62 & 14        					& $9.13^{+0.11}_{-0.10}$      	& \multicolumn2c{--}         			& $9.30^{+0.31}_{-0.31}$      			& 9.53 & 0.09						& 1,7,10,11\\
... & \multicolumn2c{...} & ... & \multicolumn2c{...} & ... & \multicolumn2c{...} & ... \\
2095			& 249 & 14      					& $10.51^{+0.13}_{-0.10}$      	& \multicolumn2c{--}         			& $11.40^{+0.05}_{-0.04}$       		& 11.45 & 0.03						& 1,2,4 \\
\hline\hline
\end{tabular}
\caption{SHIVir Sample Masses. This table lists circular velocities and masses for the 138 Virgo Cluster galaxies with raw spectra: (1) VCC designation, (2) Circular velocity, measured as the weighted average of the values computed from the literature sources listed in the last column, (3) Stellar mass, (4) HI gas mass from ALFALFA $\alpha$.100 catalogue \citep{Haynes2011}, (5) Dark matter mass (note VCC 1188, VCC 1675, and VCC 2006 have more stellar than dynamical mass, resulting in no dark matter mass), (6) Dynamical mass within $R_{23.5}$, (7) Literature sources for kinematics (1: This work, 2: ACSVCS, 3: ALFALFA \citep{Haynes2011}, 4: ATLAS$^{\rm 3D}$ \citep{Cappellari2011a}, 5: \citet{Chemin2006}, 6: \citet{Fouque1990}, 7: \citet{Geha2003}, 8: \citet{Rubin1997} and \citet{Rubin1999}, 9: \citet{Rys2014}, 10: SMAKCED \citep{Toloba2011}, 11: \citet{vanZee2004}).}
\label{tab:masses}
\end{center}
\end{table*}



\bibliographystyle{mnras}
\bibliography{SHIVir_DataPaper_arXiv} 





\bsp	
\label{lastpage}
\end{document}